\documentclass[letterpaper,twocolumn,10pt,anonymous=true]{article}
\usepackage{usenix-2020-09}

\usepackage{url}
\usepackage{graphicx}
\usepackage{tikz}
\usetikzlibrary{arrows.meta,calc,positioning,shapes.geometric}
\usepackage{amsmath}
\usepackage{bbold}
\usepackage{booktabs}

\usepackage[utf8]{inputenc}
\usepackage[ruled,vlined,linesnumbered,noend]{algorithm2e}
\usepackage{subcaption}
\usepackage{pifont}
\usepackage{xspace}
\usepackage{breakurl}
\usepackage{listings}
\usepackage{multirow}
\usepackage{makecell}
\usepackage{framed}
\usepackage{xcolor}

\definecolor{myleftbarcolor}{rgb}{0.14, 0.22, 0.62}
\definecolor{myshadecolor}{rgb}{0.941, 0.937, 0.996}

\newenvironment{coloredleftbar}{%
  \MakeFramed{\advance\hsize-\width \FrameRestore}
}{%
  \endMakeFramed %
}

\usepackage[normalem]{ulem}

\AtBeginDocument{%
  }
    
\definecolor{ForestGreen}{rgb}{0.13, 0.55, 0.13} 
\definecolor{burgundy}{rgb}{0.5, 0.0, 0.13}

\graphicspath{{./figures/}}

\SetCommentSty{mycomment}
\AddToHook{cmd/@algocf@start/after}{%
  \setlength{\hsize}{\dimexpr\linewidth-\algomargin\relax}}
\let\oldnl\nl%
\newcommand{\nonl}{\renewcommand{\nl}{\let\nl\oldnl}}
\newcommand{\algoroleline}[3]{\nonl\textcolor{#1}{\scalebox{1.2}{$\triangleright$} #2} #3}

\newcommand{\sysname}{\textsc{BFTide}\xspace} %

\usepackage{enumitem}

\newlist{todoitems}{itemize}{1}
\setlist[todoitems]{leftmargin=1.6em, itemsep=0pt, topsep=5pt}

\begin{document}

\title{Adaptive Switching Between Leader-Based and Leaderless BFT Protocols}

\author{
{\rm Sudip Bhujel} \\ University of Kentucky
\and {\rm Yue Li} \\ University of Kentucky
\and {\rm Ning Zhang} \\ Washington University in St. Louis
\and {\rm Y. Thomas Hou} \\ Virginia Tech
\and {\rm Wenjing Lou} \\ Virginia Tech
\and {\rm Yang Xiao} \\ University of Kentucky
}

\maketitle
\pagestyle{plain}
\begin{abstract}

Byzantine fault-tolerant (BFT) protocols are known for providing operational consistency and resilience in distributed systems. However, evolving network conditions, often driven by the network's inherent dynamism or adversarial influence, make it suboptimal to rely on a static protocol at all times. Existing BFT protocol adaptation solutions switch only among leader-based protocols and coordinate each switch through a separate consensus round, leaving them ineffective at handling severe asynchrony or situations in which an adaptive adversary targets the network's leader.

We propose \textsc{BFTide}, a protocol adaptation architecture that enables a BFT system to intelligently and swiftly switch to a suitable protocol as network conditions shift. \textsc{BFTide} integrates a novel protocol switching layer that embeds protocol transition logic into the ongoing BFT operation, enabling safe and low-overhead transitions between partially synchronous leader-based protocols and asynchronous leaderless protocols. It further incorporates an offline-trained reinforcement learning policy that allows nodes to propose protocols at runtime based on observed system metrics. Experimental results show that \textsc{BFTide} reduces transaction latency under adverse network conditions compared with static BFT protocols and the state-of-the-art BFT protocol adaptation scheme BFTBrain (NSDI'25), while maintaining comparable throughput. The switching layer adds a modest 10--21\% overhead to median latency when idle and requires no separate consensus round per switch.

\end{abstract}

\section{Introduction}
\label{sec:intro}

Byzantine fault-tolerant (BFT) consensus ensures that non-faulty participants (nodes) agree on a common value despite arbitrarily behaving (Byzantine) participants. BFT state machine replication (BFT-SMR) \cite{schneider1990implementing} extends BFT consensus to transaction processing by ensuring all non-faulty nodes agree on each transaction's outcome and execute transactions in the same order \cite{defago2004total}.
Consequently, BFT-SMR is widely used in database replication \cite{clement2009upright,corbett2013spanner,suri2021basil}, distributed ledgers, and fault-tolerant computing systems \cite{kwon2014tendermint,androulaki2018hyperledger,baudet2019state,lovejoy2025heraclius}.

The communication redundancy of BFT-SMR protocols (BFT protocols for short) makes their performance sensitive to network conditions.
Past research has shown that no single BFT protocol fits all network scenarios because protocols differ in their design priorities and constraints~\cite{cachin2017blockchain,xiao2020survey,amiri2024bedrock}.
For instance, PBFT \cite{castro1999practical} and its variants \cite{abd2005fault,cowling2006hq,martin2006fast,kotla2007zyzzyva,clement2009making,yin2019hotstuff} adopt the leader-based, pipelined paradigm along with timeout mechanisms to achieve high throughput during favorable (synchronous) network conditions.
However, they suffer from frequent protocol view changes and prolonged latency when the network connectivity fluctuates dramatically, and when an adversary targets the leader, they can degrade to latencies several times those of a leaderless protocol, or stall \cite{gelashvili2022jolteon}. 
In contrast, asynchronous BFT protocols \cite{rabin1983randomized,ben1993asynchronous,miller2016honey,duan2018beat,abraham2019asymptotically,guo2020dumbo,liu2020epic,zhang2023waterbear,duan2023fin} are predominantly based on a leaderless design in that all nodes independently accept client requests and drive protocol progress. They make no timing assumptions on message delivery and are not impacted by single-node bottlenecks. 
However, they rely on expensive cryptographic primitives to guarantee the protocol's eventual termination (i.e., liveness), resulting in significant computation and communication overhead \cite{gelashvili2022jolteon}, which is undesirable when the network becomes synchronous \cite{gelashvili2021prepared}.

\vspace{3pt}
\noindent
\textbf{Adapting to Evolving Network Conditions.}
Changing network conditions and adversarial influence make static protocol configurations suboptimal for both latency and throughput \cite{berger2020aware,gramoli2023diablo,wu2025bftbrain}. Prior work explores two approaches to improving real-time performance.

The \emph{dual-path} approach combines optimistic and pessimistic paths for network adaptability \cite{blum2019synchronous,neu2021ebb,crain2021red,gelashvili2021prepared,gelashvili2022jolteon,blum2023abraxas}. The optimistic path usually runs a partially synchronous protocol for high performance under favorable conditions. The pessimistic path usually runs an asynchronous BFT protocol to preserve safety and eventual liveness under extreme network conditions.
The dual-path approach requires each participant to decide which path to use for client requests and result commitment
based on a pre-defined trigger.
The trigger, however, requires frequent and careful tuning, which is inefficient and not always optimal. 
The parallel variants \cite{neu2021ebb,blum2023abraxas,dai2023parbft} require running two paths simultaneously, which is inefficient if pessimistic conditions occur only rarely.

The second approach leverages \emph{multi-protocol switching}, in which system nodes proactively switch to a new protocol from a candidate pool to improve performance across a time window amid varying network conditions \cite{guerraoui2010next,bahsoun2015making,wu2025bftbrain}. In particular, BFTBrain~\cite{wu2025bftbrain} uses reinforcement learning (RL) to select a throughput-maximizing protocol from a portfolio of PBFT-style leader-based protocols based on network interactions. BFTBrain and prior multi-protocol switching systems~\cite{guerraoui2010next,bahsoun2015making,wu2025bftbrain} primarily target synchronous or partially synchronous protocols and lack a clear strategy for pessimistic asynchronous conditions. They also do not address adaptive leader manipulation: leader-based protocols such as HotStuff \cite{yin2019hotstuff} concentrate progress in the current leader and rely on view changes for leader replacement, so targeted delays or denial-of-service attacks on successive leaders can trigger repeated view changes and harm liveness during asynchronous periods.

Switching between these protocol families requires reconciling their distinct commit structures; for instance, HotStuff commits a chain of blocks, whereas asynchronous protocols rely on agreement on subsets of proposals. Crucially, replicas must agree on both the target protocol and the boundary at which it takes over, preserve the committed prefix, and retain pending requests for re-proposal. Protocol selection must therefore account for the cost of this handoff, while handoff execution remains subject to the incumbent protocol's progress.

\begin{coloredleftbar}
    \noindent\textbf{Research Question.} \textit{How can a BFT-SMR system, while its incumbent protocol continues to commit, adapt between protocols with different synchrony assumptions while preserving a common committed log and keeping transition overhead low?}
\end{coloredleftbar}

\vspace{3pt}
\noindent
\textbf{Contributions.}
We propose \sysname, a BFT protocol adaptation architecture for switching between leader-based partially synchronous and leaderless asynchronous protocols under changing network conditions without interrupting continuous operation.
As shown in Fig.~\ref{fig:switchlayer}, \sysname separates the coordination of a transition from the policy that proposes when and where to switch. It comprises two complementary components:

\begin{figure}[t]
    \centering
    \includegraphics[width=0.67\linewidth]{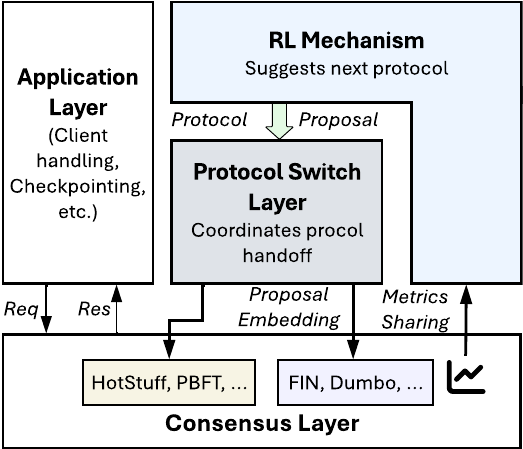}
    \caption{Overview of \sysname. The policy proposes a protocol, while the switch layer coordinates its activation.}
    \label{fig:switchlayer}
\end{figure}

\textbf{Embedded protocol switching logic.} \sysname's \emph{protocol switch layer} coordinates the target protocol and transition boundary so that the target continues the incumbent's committed log.
The design \emph{embeds} switch agreement in the incumbent's native messaging and consensus flow. Signed votes form a \emph{switch certificate} for the target. Replicas verify this evidence and reach the agreed boundary before activating the target. Under sustained application traffic, embedding allows switch agreement to share the communication and ordering costs of ordinary SMR operations. Switching progress remains subject to the incumbent's liveness conditions; safety of the protocol transition holds under the Byzantine model plus one delivery condition on the certificate. (see \S\ref{sec:base-system}.)

\textbf{Policy-guided protocol proposals.} We formulate protocol selection as a Markov Decision Process (MDP) and instantiate the policy with a Deep Q-Network (DQN). Aggregated measurements of transaction latency, throughput, offered load, and peer delays inform protocol proposals, while the reward accounts for performance and switching cost. The system loads an offline-trained model for local inference at each replica. The policy proposes a target, whereas the PS Layer governs activation through certificate verification and the agreed boundary. This separation allows other selection policies to use the same switching mechanism. (see \S\ref{sec:rl}.)

\textbf{Implementation and Validation.} 
We implemented and open-sourced \sysname along with network functions that simulate varying adverse network conditions. We deployed our experimental network on CloudLab \cite{Duplyakin+:ATC19} with up to 31 nodes. Our experiments demonstrate that \sysname can reduce latency relative to BFTBrain \cite{wu2025bftbrain} under leader delay and global delay, while achieving performance comparable to HotStuff under benign conditions. 
Notably, with a 250 ms injected leader delay in the 31-node system, \sysname tracks the better-performing protocol under each network condition. At the highest load (120 KB/s), it achieves a median latency of 3.9 s, compared with 6.3 s for HotStuff and 26.4 s for BFTBrain, while matching HotStuff's latency under benign, jitter, and mild global-delay conditions. Averaged across the four conditions at 120 KB/s, its latency is about 70\% lower than BFTBrain's.
(see \S\ref{sec:eval}.)

\section{Background and Problem Description}
\label{sec:background}

\subsection{BFT Protocols}
\label{subsec:background-bft-protocols}

A BFT-SMR protocol (``BFT protocol'' for short) extends BFT consensus into a transaction processing system. Under the assumption that $f$ out of $n$ nodes are faulty, a BFT protocol needs to satisfy three requirements \cite{schneider1990implementing,attiya2004distributed}: \emph{(i) agreement}---all non-faulty nodes decide on the same transactions for execution; \emph{(ii) total order}---all non-faulty nodes execute the same sequence of operations in the same order; and \emph{(iii) liveness}---all client transactions will eventually be executed and committed by the non-faulty nodes. Agreement and total order are often jointly known as \emph{safety}.

Classified by the network synchrony assumption, existing practical BFT protocols fall into two categories---partially synchronous and asynchronous. Partially synchronous protocols, such as PBFT \cite{castro1999practical} and its variants \cite{abd2005fault,cowling2006hq,martin2006fast,kotla2007zyzzyva,clement2009making,kwon2014tendermint,yin2019hotstuff}, designate a \emph{leader} node to take in client transactions and start a protocol session. They excel in the \emph{synchronous} condition that message delivery is guaranteed within a fixed delay $\Delta$ between any two nodes, or the \emph{partially synchronous} condition that the $\Delta$ bound holds after an unknown global stabilization time (GST) \cite{dwork1988consensus}. These protocols are usually accompanied by a view-change protocol that allows non-leader nodes (replicas) to vote out unresponsive or misbehaving leaders and elect a new leader.

In contrast, asynchronous BFT protocols \cite{rabin1983randomized,ben1993asynchronous,miller2016honey,duan2018beat,abraham2019asymptotically,guo2020dumbo,liu2020epic,zhang2023waterbear,duan2023fin} accommodate the \emph{asynchronous} condition that there is no time guarantee (nor a globally synchronized clock) on messaging, except for eventual delivery. These protocols follow the \emph{leaderless} design, where all nodes accept client transactions and equally contribute to protocol progress, typically aiming to achieve the \emph{asynchronous common subset (ACS)} consensus goal, i.e., all nodes finalize a common subset of client transactions proposed across the system. ACS-based asynchronous BFT protocols ensure safety and eventual liveness, not subject to any timing constraints or single-node bottlenecks. They, however, rely on complex communication and cryptographic mechanisms, such as threshold signatures \cite{shoup2000practical} and common coins \cite{canetti1993fast}, to achieve such liveness.

\vspace{2pt}
\noindent
\textbf{Representative Protocols.} We select the PBFT variants as representative partially synchronous, leader-based protocols; we select Asynchronous Common Subset (ACS) as representative asynchronous, leaderless protocols.
Their protocol schematics are shown in Appendix \ref{app:bft-protocols}.

\textbf{PBFT}~\cite{castro1999practical} is a classic leader-based BFT protocol that progresses through a sequence of views, each coordinated by a primary. The main protocol has three phases---\textit{pre-prepare}, \textit{prepare}, and \textit{commit}---which ensure all non-faulty replicas agree on the ordering of client requests. In the pre-prepare phase, the leader assigns a sequence number; the prepare phase guarantees its uniqueness within the view; and the commit phase ensures safety across views, allowing recovery after leader failures. Clients send requests to the leader, which orchestrates the three phases via multicasts and waits for $2f{+}1$ matching messages before commitment and execution.

\textbf{HotStuff}~\cite{yin2019hotstuff} is a PBFT-like protocol with linear communication complexity. 
Each consensus round has three phases: \textit{prepare}, \textit{precommit}, and \textit{commit}. Unlike PBFT, the leader in HotStuff sends a message to all replicas in each phase, and the replicas respond with votes. Once the leader collects $2f{+}1$ votes, it forms a Quorum Certificate (QC), which is used to advance to the next phase.
HotStuff's consensus progress relies on the leader's responsiveness. Under high delay or adversarial conditions targeting the leader, HotStuff can experience significant performance degradation due to frequent view changes and the leader bottleneck.

\textbf{ACS} is a general type of asynchronous BFT protocol that is based on a leaderless design~\cite{ben1993asynchronous,miller2016honey,duan2018beat,guo2020dumbo,duan2023fin}. 
ACS tolerates fully asynchronous and arbitrary peer-wise message delays, making it resilient to network partitions and adaptive attacks.
The general ACS protocol proceeds in two phases. In the first phase, each node runs a Reliable Broadcast (RBC) instance to disseminate its input. RBC consists of three substeps---\textit{VAL}, \textit{ECHO}, and \textit{READY}---which together guarantee that all non-faulty nodes eventually agree on whether and what to accept from a given sender. Once all RBCs are complete, the second phase begins: a Byzantine Agreement (BA) protocol is used to collectively decide on a common subset of accepted inputs across all nodes. Extending from the general ACS framework, \textbf{Dumbo} \cite{guo2020dumbo} replaces the parallel BAs with a multi‑valued validated Byzantine agreement (MVBA) to reduce communication complexity; \textbf{FIN} \cite{duan2023fin} relies on a novel signature-free MVBA design to achieve $O(1)$ time complexity, which, according to our experiments, achieves the best consensus latency performance.

\subsection{Challenges for BFT Protocol Switching}
\label{subsec:background-challenges}

To effectuate performance-optimized BFT protocol switching for a changing network condition,
we face three main design challenges.

\colorbox{gray!20}{\emph{Challenge-1:}} The execution of protocol switching requires agreement from all nodes, which in itself constitutes a BFT consensus problem. 
This ``meta-consensus'' should be swift and minimally impact the ongoing SMR operation.

\colorbox{gray!20}{\emph{Challenge-2:}} The transition between a partially synchronous leader-based protocol and an asynchronous leaderless protocol is non-trivial since they realize BFT-SMR differently. For instance, ACS schemes require every node to accept client requests and use a canonical order to execute the agreed subset of requests, whereas in PBFT variants, the leader monopolizes request handling and can impose an arbitrary order on executing them. %
Crucially, the transitions between different protocols should not break the safety or liveness property of the ongoing BFT-SMR operation.

\colorbox{gray!20}{\emph{Challenge-3:}} The decision to change protocols should lead to optimal performance for the current network condition.
Each node should be able to propose when and what to switch to based on individual observations of the network status, before contributing to the meta-consensus (\colorbox{gray!20}{\emph{Challenge-1}}). Making an optimal switching decision is ultimately dependent on intelligence gathering by each node. 

\section{The \sysname Architecture}
\label{sec:base-system}

The overall system architecture of \sysname is illustrated in Fig.~\ref{fig:switchlayer}. To support protocol adaptation in decentralized systems with minimal overhead, we introduce two components: the \emph{Protocol Switch (PS) Layer} and the \emph{RL mechanism}. 

The PS Layer functions as a control plane for finalizing protocol selection, ensuring all decentralized agents agree on the same protocol for the next epoch efficiently and safely. This layer abstracts protocol switching enforcement away from the ongoing consensus layer (which runs the incumbent protocol) and protocol proposal generation (in the RL mechanism). In this section, we focus on the PS Layer functions and defer the RL mechanism to the next section.

\subsection{Network Model}
\label{subsec:network-model}

We consider a static set of $n=3f+1$ replicas connected in a decentralized network, where each pair communicates over an authenticated channel. Up to $f$ replicas may be Byzantine and behave arbitrarily; the remaining replicas, which we call \emph{honest}, follow the protocol faithfully. Membership and cryptographic identities remain fixed during an execution. We assume that signatures cannot be forged and that honest replicas preserve their durable protocol state across restarts. Clients follow the request protocol and may retry requests. HotStuff and FIN tolerate the same $f$ with $2f{+}1$ quorums, so a fault placement that one tolerates, the other tolerates as well.

Network conditions vary in peer-wise message delay and bandwidth. Under partial synchrony, messages between honest replicas are delivered within a bound $\Delta$ after an unknown stabilization time; this is the timing condition required for HotStuff's liveness. Under asynchrony, message delays have no fixed upper bound. We assume the \emph{reliable channel} property: a message that an honest replica retransmits indefinitely is eventually delivered to its honest recipient. Our prototype obtains retransmission from TCP and recorded no dropped messages in the reported runs. FIN relies on eventual delivery and its randomized agreement assumptions for liveness, without requiring a fixed delay bound. Safety must hold despite message delays; progress remains subject to the active protocol's liveness assumptions.

The network adversary may observe the current leader and public configuration, delay, duplicate, or reorder messages on any subset of links, and vary link bandwidth. Finite message loss is permitted subject to eventual delivery under retransmission. These network actions do not increase the number of corrupted replicas: the bound of $f$ applies to replica corruption, while network interference may affect links incident to any replica. The adversary cannot forge an honest replica's identity or signature. A persistent partition that prevents communication by a quorum is outside the liveness conditions, although safety must still hold.

Switching itself also requires incumbent progress. A switch certificate names a boundary height in the incumbent's log, and the incumbent must commit through that height before the target can take over. Consequently, \sysname can adapt under degraded conditions while the incumbent continues to make progress, but cannot guarantee escape from an incumbent that has stopped committing. A local timeout or an RL proposal alone does not authorize activation of another protocol. This distinction separates safe adaptation from recovery under a persistent loss of quorum communication.

\subsection{Protocol Switch Layer Functions}
\label{subsec:base-system}

Protocol switching requires agreement on both the target and where it takes over in the application log. In the \emph{embedded design}, replicas directly exchange signed switch votes, and the certificate identifies the incumbent-log height at which the target takes over. Activation waits for incumbent commitment through that height, without a separate consensus round, avoiding BFTBrain's per-decision cost.

\begin{algorithm}[t]
    \caption{\sysname Selection (for node $i$)}
    \label{alg:hotswitch}

    \small

    \tcc{State: $incumbent$, window $w$, fault bound $f$; $D_s$: received valid votes for the window ending at $s$.}

    \For{$t \gets 1, 2, \ldots$}{
        \If{$t > 2w$ \textbf{and} $(t-1) \bmod w = 0$}{
            $s \gets t-1-w$;\quad $\bar{m}^s \gets \textsc{AgreedMetrics}(\text{reports for } s)$\;
            $h^s \gets H(\bar{m}^s)$;\quad $\hat{a} \gets \mathrm{policy}(\bar{m}^s)$;\quad $v_i^s \gets \bot$\;
            \If{$\hat{a} \neq incumbent$ \textbf{and} dwell satisfied}{
                $b_s \gets s + (k+1)\cdot w$;\quad $\sigma_i \gets \textsf{Sign}_i(H(s \,\Vert\, \hat{a} \,\Vert\, i \,\Vert\, h^s \,\Vert\, b_s))$\;
                $v_i^s \gets \langle s,\,\hat{a},\,i,\,h^s,\,b_s,\,\sigma_i \rangle$\tcp*{One vote per window.}
            }

            \algoroleline{orange}{As a replica}{\tcp{Vote dissemination.}}
            \If{$v_i^s \neq \bot$}{
                attach $v_i^s$ to next incumbent message per peer\;
            }
            add received valid votes to $D_s$\;

            \algoroleline{orange}{Every replica, on $|D_s(a)| \geq 2f{+}1$}{}{\DontPrintSemicolon\;}
            $a_s \gets \mathrm{EvaluateSwitch}(s, \textsf{votes\_in}(D_s))$\;
        }

        \algoroleline{orange}{End of round $t$}{}{\DontPrintSemicolon\;}
        record $\hat{m}_t$\;
        \If{$a_s \neq incumbent$ \textbf{and} log committed through $b_s$}{
            activate $a_s$\;
        }
    }

    \SetKwFunction{FEval}{EvaluateSwitch}
    \SetKwProg{Fn}{Function}{}{end}
    \Fn{\FEval{$s$, $votes$}}{
        $A \gets \{v \in votes : v.\textit{window}=s,\ v.\textit{target} \neq incumbent,\ v.h = h^s,\ v.b = b_s,$\newline
        \hspace*{1em}$v$ has a valid sender signature$\}$\;
        \If{$\exists a : |\{v.sender : v \in A,\ v.\textit{target}=a\}| \geq 2f{+}1$}{
            \Return{$a$}\;
        }
        \Return{$incumbent$}\tcp*{Retain incumbent.}
    }
\end{algorithm}

\begin{coloredleftbar}
    \noindent\textbf{Key idea.} \textit{Replicas derive the same target from a $2f{+}1$ certificate whose votes are bound to the same metric digest and the same boundary height. The target continues from the application state at that height of the incumbent's log.}
\end{coloredleftbar}

Algorithm~\ref{alg:hotswitch} sketches target selection in the embedded design,
abstracting their regular message rounds. Its output is a selected protocol; the handoff requirements below govern when that protocol becomes the incumbent.

\subsubsection{Windows, Metrics, and Switch Votes}
Each node monitors the consensus execution. At the end of each round $t$, node $i$ records local metrics $\hat{m}_t$ (to elaborate in \S\ref{subsec:mdp}). After each $w$-round \emph{synchronization window} (hereafter, \emph{window}), it attaches an authenticated report to the incumbent's next message to each peer and aggregates at least $2f{+}1$ reports into agreed metrics $\bar{m}$ (\S\ref{metrics-synchronization}). The policy uses the previous window's aggregate, allowing one window for report arrival. A node votes only if the policy proposes the same target in two consecutive windows and at least five windows have elapsed since the last switch. This dwell period doubles after a rapid re-switch to prevent thrashing (\S\ref{sec:sensitivity}). Without an accepted switch, the system retains the incumbent.

The voting window determines when replicas consider a switch. The certified switch decision identifies where the target takes over in the application log. The following description considers the window ending at round $w$; the same rules apply to subsequent windows.

At the close of the window that follows $w$, once the reports for $w$ have arrived, each node emits at most one signed switch vote $v_i^w$ for a target protocol $a$: 
\[
    v_i^w = \langle w,\, a,\, sender\_id=i,\, H(\bar{m}),\, b_w,\, \sigma_i \rangle
\]
where $\bar{m}$ is computed by \textsc{AgreedMetrics} in Algorithm~\ref{alg:metrics-sync}, $H(\bar{m})$ is its digest, $b_w = w + (k+1)\cdot w'$ is the boundary height, $k$ windows of $w'$ rounds after the voting window for vote dissemination and certificate assembly ($k{=}3$ in our deployment), and $\sigma_i=\textsf{Sign}_i(H(w\,\|\,a\,\|\,i\,\|\,H(\bar{m})\,\|\,b_w))$. An honest node never signs two distinct votes for the same window. Votes travel to all replicas the way reports do, on the incumbent's next message to each peer.

We use $D_w$ to denote the set of valid votes for the window ending at round $w$ that a replica has received. Because each vote carries $H(\bar{m})$, votes with different metric digests do not combine: $\textsf{votes\_in}(D_w)$ counts only votes that agree on the target, the metric digest, and the boundary.

\subsubsection{Switch Certificate} 
A switch certificate $C_w(a)$ forms when at least $2f{+}1$ distinct signed votes for protocol $a$ appear in $D_w$. For deterministic verification, each honest node selects a canonical subset of $2f{+}1$ eligible votes, using the smallest $sender\_id$ values, and a node that assembles the certificate forwards it, so a replica whose own copy of $D_w$ is incomplete still learns the decision.
Without a valid $C_w(a)$, every honest node retains the incumbent. Because honest nodes cast at most one vote per window, quorum intersection ensures that at most one target can satisfy the $2f{+}1$ threshold, whatever the faulty nodes sign. Each vote names its target, so certifying a specific protocol needs no multi-valued agreement.

\subsubsection{Handoff}
A certificate $C_w(a)$ fixes the target and terminal height $b_w$. It need not carry the prefix. Incumbent safety and deterministic execution give replicas committed through $b_w$ the same application state and per-client deduplication watermarks.

\emph{Stopping.} Certificate holders complete the incumbent prefix through $b_w$ and neither support nor install incumbent decisions above it. A sufficient condition is that at least $f{+}1$ honest replicas receive $C_w(a)$ before sending any votes or epoch messages supporting decisions above $b_w$. This leaves at most $2f$ participants, preventing a $2f{+}1$ quorum. Later receipt cannot revoke earlier votes. The $k$-window lead, 15 rounds for $k{=}3$, gives the piggybacked votes and certificate time to reach every replica. Timely delivery remains an additional safety premise, which neither partial synchrony nor FIN's eventual delivery guarantees for fixed $k$.

\emph{Activation.} Activation requires a verified $C_w(a)$ and the committed prefix through exactly $b_w$. Timeouts waive neither condition. HotStuff restarts at view~1 of its configured leader schedule, anchored at $b_w$, and FIN starts at epoch $b_w{+}1$. Messages for the incoming instance are buffered until activation, and messages from obsolete instances are discarded. Catch-up installs consecutive heights through $b_w$, checking each height's payloads against matching committed-data digests from $f{+}1$ distinct authenticated peers. At least one is honest, so incumbent agreement determines the content.

\emph{Completion.} The handoff completes locally at activation. With $n{-}f$ honest replicas in the same instance, the target eventually progresses under its liveness assumptions. Later replicas use verified catch-up while the required data remain available. Locally retained uncommitted requests are re-queued with their identities intact, and inherited watermarks prevent repeated execution. Re-proposal may repeat speculative work and does not guarantee inclusion in the first target decision. Fig.~\ref{fig:skew} reports activation lag.

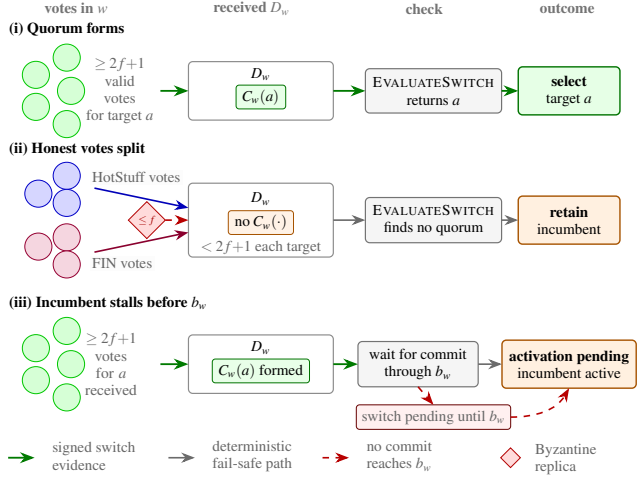
\begin{figure}
    \centering
    \resizebox{\columnwidth}{!}{\begin{tikzpicture}[
    every node/.style={outer sep=0pt},
    font=\scriptsize,
    >=Stealth,
    panel/.style={
        draw=black!18,
        fill=black!1,
        rounded corners=3pt,
        line width=.45pt,
        inner sep=5pt
      },
    stage/.style={
        font=\scriptsize\bfseries,
        text=black!55,
        align=center
      },
    replica/.style={
        circle,
        draw=#1!78!black,
        fill=#1!16,
        line width=.45pt,
        minimum size=4.4mm,
        inner sep=0pt
      },
    byz/.style={
        diamond,
        aspect=1,
        draw=red!70!black,
        fill=red!14,
        line width=.45pt,
        minimum size=5.6mm,
        inner sep=0pt
      },
    artifact/.style={
        rectangle,
        rounded corners=2pt,
        draw=black!45,
        fill=white,
        line width=.55pt,
        minimum width=2.35cm,
        minimum height=.95cm,
        align=center
      },
    eval/.style={
        rectangle,
        rounded corners=2pt,
        draw=black!55,
        fill=black!4,
        line width=.55pt,
        minimum width=1.95cm,
        minimum height=.72cm,
        align=center
      },
    switch/.style={
        rectangle,
        rounded corners=2pt,
        draw=green!45!black,
        fill=green!10,
        line width=.65pt,
        minimum width=1.65cm,
        minimum height=.78cm,
        align=center
      },
    retain/.style={
        rectangle,
        rounded corners=2pt,
        draw=orange!65!black,
        fill=orange!12,
        line width=.65pt,
        minimum width=1.65cm,
        minimum height=.78cm,
        align=center
      },
    cert/.style={
        rectangle,
        rounded corners=1.4pt,
        draw=green!45!black,
        fill=green!9,
        inner xsep=3pt,
        inner ysep=2pt,
        align=center
      },
    nocert/.style={
        rectangle,
        rounded corners=1.4pt,
        draw=orange!65!black,
        fill=orange!10,
        inner xsep=3pt,
        inner ysep=2pt,
        align=center
      },
    badgate/.style={
        rectangle,
        draw=red!72!black,
        fill=red!10,
        line width=.75pt,
        minimum width=.26cm,
        minimum height=1.28cm
      },
    flow/.style={->, line width=.75pt, draw=black!58},
    goodflow/.style={->, line width=.85pt, draw=green!48!black},
    splitflow/.style={->, line width=.75pt, draw=#1!72!black},
    badflow/.style={->, dashed, line width=.75pt, draw=red!72!black},
    blockflow/.style={-, dashed, line width=.75pt, draw=red!72!black},
    note/.style={font=\scriptsize, text=black!60, align=center}
  ]

  \foreach \x/\txt in {
  1.05/{votes in $w$},
  3.90/{received $D_w$},
  6.70/{check},
  9.05/{outcome}
  } {
  \node[stage] at (\x,1.22) {\txt};
  }

  \begin{scope}[shift={(0,0)}]
    \node[font=\bfseries\scriptsize, anchor=west] (t1) at (-.15,.88)
    {(i) Quorum forms};

    \begin{scope}[shift={(0,-.12)}]
      \node[replica=green] (q1) at (.40,.26) {};
      \node[replica=green] (q2) at (.90,.54) {};
      \node[replica=green] (q3) at (.97,0) {};
      \node[replica=green] (q4) at (.90,-.54) {};
      \node[replica=green] (q5) at (.40,-.26) {};
      \node[note, anchor=west] at (1.10,0)
      {$\ge 2f{+}1$\\valid\\votes\\for target $a$};

      \node[artifact] (d1) at (4.05,0) {};
      \node[font=\scriptsize\bfseries] at (4.05,.27) {$D_w$};
      \node[cert] at (4.05,-.10) {$C_w(a)$};

      \node[eval] (e1) at (6.85,0) {\textsc{EvaluateSwitch}\\returns $a$};
      \node[switch] (o1) at (9.05,0) {\textbf{select}\\target $a$};

      \draw[goodflow] (2.42,0) -- (d1.west);
      \draw[goodflow] (d1.east) -- (e1.west);
      \draw[goodflow] (e1.east) -- (o1.west);
    \end{scope}
  \end{scope}

  \begin{scope}[shift={(0,-2.22)}]
    \node[font=\bfseries\scriptsize, anchor=west] (t2) at (-.15,1.15)
    {(ii) Honest votes split};

    \node[replica=blue] (h1) at (.44,.43) {};
    \node[replica=blue] (h2) at (.90,.72) {};
    \node[replica=blue] (h3) at (.90,.28) {};
    \node[replica=purple] (f1) at (.44,-.43) {};
    \node[replica=purple] (f2) at (.90,-.28) {};
    \node[replica=purple] (f3) at (.90,-.72) {};
    \node[note, fill=black!1, inner sep=1pt, anchor=west] at (1.28,.70) {HotStuff votes};
    \node[note, fill=black!1, inner sep=1pt, anchor=west] at (1.28,-.70) {FIN votes};
    \node[byz, minimum size=6mm, font=\tiny, text=red!65!black] (bz)
    at (2.20,0) {$\le f$};

    \node[artifact, minimum height=1.20cm] (d2) at (4.05,0) {};
    \node[font=\scriptsize\bfseries] at (4.05,.35) {$D_w$};
    \node[nocert] at (4.05,-.04) {no $C_w(\cdot)$};
    \node[note] at (4.05,-.45) {$<2f{+}1$ each target};

    \node[eval] (e2) at (6.85,0) {\textsc{EvaluateSwitch}\\finds no quorum};
    \node[retain] (o2) at (9.05,0) {\textbf{retain}\\incumbent};

    \draw[splitflow=blue] (1.35,.52) -- ($(d2.west)+(0,.20)$);
    \draw[splitflow=purple] (1.35,-.52) -- ($(d2.west)+(0,-.20)$);
    \draw[badflow] (bz.east) -- (d2.west);
    \draw[flow] (d2.east) -- (e2.west);
    \draw[flow] (e2.east) -- (o2.west);
  \end{scope}

  \begin{scope}[shift={(0,-4.44)}]
    \node[font=\bfseries\scriptsize, anchor=west] (t3) at (-.15,.88)
    {(iii) Incumbent stalls before $b_w$};

    \begin{scope}[shift={(0,-.12)}]
      \node[replica=green] (s1) at (.40,.26) {};
      \node[replica=green] (s2) at (.90,.54) {};
      \node[replica=green] (s3) at (.97,0) {};
      \node[replica=green] (s4) at (.90,-.54) {};
      \node[replica=green] (s5) at (.40,-.26) {};
      \node[note, text width=.95cm] at (1.62,0)
      {$\ge 2f{+}1$ votes\\for $a$\\received};

      \node[artifact] (d3) at (4.05,0) {};
      \node[font=\scriptsize\bfseries] at (4.05,.27) {$D_w$};
      \node[cert] at (4.05,-.11) {$C_w(a)$ formed};

      \node[eval] (e3) at (6.6,0) {wait for commit\\through $b_w$};
      \node[retain] (o3) at (9.05,0) {\textbf{activation pending}\\incumbent active};
      \node[note, draw=red!40!black, fill=red!6, rounded corners=1.5pt, inner sep=2.5pt] (vc)
      at (6.85,-.86) {switch pending until $b_w$};

      \draw[goodflow] (2.42,0) -- (d3.west);
      \draw[goodflow] (d3.east) -- (e3.west);
      \draw[flow] (e3.east) -- (o3.west);
      \draw[badflow] (e3.south) -- (vc.north);
      \draw[badflow] (vc.east) to[out=0,in=-120] (o3.south);
    \end{scope}
  \end{scope}

  \begin{scope}[shift={(0,-6.08)}]
    \draw[goodflow] (-.05,0) -- (.38,0);
    \node[note, anchor=west, align=left, yshift=.03cm] at (.55,0) {signed switch\\evidence};
    \draw[flow] (2.55,0) -- (2.98,0);
    \node[note, anchor=west, align=left, yshift=.03cm] at (3.15,0) {deterministic\\fail-safe path};
    \draw[badflow] (5.05,0) -- (5.48,0);
    \node[note, anchor=west, align=left, yshift=.03cm] at (5.65,0) {no commit\\reaches $b_w$};
    \node[byz, minimum size=3.2mm] at (8.10,0) {};
    \node[note, anchor=west, align=left, yshift=-.02cm] at (8.38,0) {Byzantine\\replica};
  \end{scope}

\end{tikzpicture}}
    \vspace{-1.5em}
    \caption{Three switch-vote outcomes at window $w$.}
    \label{fig:voting-scenarios} 
\end{figure}

\textbf{Three illustrative scenarios.}
Fig.~\ref{fig:voting-scenarios} shows the three switch-vote outcomes at window~$w$.
\emph{(i) Quorum forms}: $D_w$ contains $\geq 2f{+}1$ valid votes for a common target $a$, so the certificate $C_w(a)$ forms and replicas select $a$ and activate it after satisfying the handoff requirements.
\emph{(ii) Honest votes split}: if no target reaches $2f{+}1$ valid votes in $D_w$, no certificate forms and the incumbent is retained.
\emph{(iii) Incumbent stalls before the boundary}: votes travel replica to replica, so a faulty leader cannot withhold them, but the certificate $C_w(a)$ takes effect only at height $b_w$. If the incumbent stops committing before $b_w$, the certificate remains pending and takes effect when the incumbent resumes committing. Because windows close upon commits, no competing certificate can form in the meantime.

These cases describe protocol selection. A replica that learns the decision later must still obtain the application state for the committed prefix before completing the handoff.

\subsubsection{Instantiation for HotStuff and FIN}

We describe how to leverage the switch certificate mechanism in two cases, switching back and forth between HotStuff (leader-based) and FIN (leaderless).

\textbf{Leader-based to Leaderless.} For a leader-based protocol such as HotStuff, each replica sends its signed switch vote $v_i^w$ to every peer on its next message and selects FIN once it holds or receives a valid $C_w(\mathrm{FIN})$. The certificate boundary $b_w$ is a block height. The replica stops voting for HotStuff blocks beyond $b_w$ and activates FIN once its log has been committed through $b_w$ under HotStuff's commit rule~\cite{yin2019hotstuff}. The leader plays no role in this decision and therefore cannot withhold it.

In HotStuff, once the block at $b_w$ commits, it has no uncommitted descendants. Under chained HotStuff, any uncommitted descendants of $b_w$ would also be excluded from the prefix. FIN starts from the application state through $b_w$. To retain pending requests, the system uses \emph{request re-queuing}: locally retained in-flight requests are made available for proposal under FIN. The ledger's per-client sequence watermarks survive the switch and suppress repeated execution. Re-proposal may repeat work spent on uncommitted branches.

\textbf{Leaderless to Leader-based.} FIN runs parallel RBC instances and uses MVBA to agree on which instances contribute to the common subset~\cite{duan2023fin}. Switch votes travel as in HotStuff, here on RBC traffic. The boundary $b_w$ is an epoch, and a replica holding a valid $C_w(\mathrm{HotStuff})$ decides all epochs through $b_w$ and discards any epoch beyond it. Otherwise, FIN remains the incumbent.

Before activating HotStuff, replicas must obtain all selected payloads and execute the terminal subset in the same deterministic order. The system orders batches by proposer ID and preserves request order within each batch. HotStuff must continue from the resulting application state and log height, using its configured leader schedule. A lagging replica therefore needs verified committed data before activation. Locally retained requests omitted from the subset remain eligible for re-proposal under HotStuff, subject to the same duplicate suppression.

\section{Learning to Select Protocols}
\label{sec:rl}

In this section, we introduce the RL mechanism that leads each node to generate switch proposals. During each \emph{window}, all nodes record their local system metrics and, at the end of the window, exchange authenticated metric reports and aggregate them to a common set of global metrics (\S\ref{metrics-synchronization}). 
A DQN is trained offline in a simulator calibrated using the latency and throughput of each candidate protocol measured on the testbed under each network condition. This calibration provides a realistic BFT environment and helps the DQN identify the optimal protocol. The calibration table has one row per protocol and condition. 

\textbf{Utilizing the Consensus Layer.} The incumbent consensus layer provides a common reference point for protocol selection. A proposal is a signed vote bound to the aggregated metrics from which it was derived, and the resulting switch certificate specifies a height in the incumbent protocol's committed log at which every replica switches protocols. Thus, the decision requires no separate consensus round.

\subsection{MDP Formulation for Protocol Selection}
\label{subsec:mdp}

We model BFT protocol selection as an MDP~\cite{suttonRLIntro2018}, where an RL agent observes the network state, proposes a protocol, and receives a performance-based reward. We adopt an MDP over standard multi-armed and contextual bandit formulations (MAB, CB)~\cite{auerUCBRevisitedImproved2010,wu2025bftbrain} to explicitly model how protocol choices affect subsequent system states and rewards. A protocol switch incurs a one-off cost, while the protocol it installs determines latency and throughput in every following window until the next switch. Discounted future rewards let the policy weigh the switching cost against the performance of the protocol it installs~\cite{kaelblingReinforcementLearningSurvey1996}.

Our MDP model is $\mathcal{M} = \langle \mathcal{S}, \mathcal{A}, \mathcal{P}, \mathcal{R}, \gamma \rangle$, where $\mathcal{S}$ is the state space, $\mathcal{A}$ the action space, $\mathcal{P}$ the transition probability distribution, $\mathcal{R}$ the reward function, and $\gamma$ the discount factor \cite{suttonRLIntro2018}. The discount factor determines the agent's preference for immediate versus future rewards.

\vspace{2pt}
\noindent
\textbf{\underline{State Space.}}
At time step $t$, the six-dimensional state $s_t = (tp_t, l_t, ld_t, d_t, p_t, a_{t-1})$ comprises three continuous performance measurements, two peer-probe-derived network indicators, and the incumbent protocol. The performance and network entries are synchronized across replicas (\S\ref{metrics-synchronization}) rather than observed locally, whereas the offered load is replica-specific. Continuous entries are normalised to deployment-specific ranges before reaching the agent.

\textbf{Throughput ($tp_{t}$):}
Throughput is the rate at which transactions are committed, measured in data finalized per second. It quantifies the system's capacity under the current protocol.

\textbf{Latency ($l_{t}$):}
Latency, defined as the request resolution time, is the duration from when a client request arrives to when the system commits it. It reflects the responsiveness of the consensus mechanism under the protocol.

\textbf{Offered Load ($ld_t$):} Offered load is the rate at which a replica's clients submit requests (KB/s). Our prototype determines this load from the configured client rate, which remains constant within each run. It lets the agent separate latency caused by the network from latency caused by queueing.

\textbf{Delayed peers ($d_t$):}
Each replica probes peers every window; $d_t$ is the fraction whose agreed round-trip time (RTT) exceeds $T$, counting nonresponses as delayed. It distinguishes a slow leader from a uniformly slow network and, under a leaderless protocol with leader-independent latency, is the only signal that the leader attack has ended.

\textbf{Incumbent protocol ($a_{t-1}$):}
The protocol currently in use. The same latency has different implications under HotStuff and FIN, and the switch penalty in the reward is meaningful only relative to the current protocol.

\vspace{2pt}
\noindent
\textbf{\underline{Action Space.}}
The discrete action space is a predefined, extensible protocol pool $Proto_L$, where $L$ denotes the unique protocol identifiers. The protocols in $Proto_L$ are either leader-based or leaderless. At step $t$, the RL agent at node $i$ selects a protocol $a_{i,t} \in Proto_L$ to propose based on the observed state.

\vspace{2pt}
\noindent
\underline{\textbf{Determining Network Status ($p_t$).}}
We derive the delay indicator from the per-peer round trips behind $d_t$. The \emph{timeout threshold} $T$ defines the maximum acceptable response delay. Node $j$ is flagged as delayed at round $t$ via the per-node boolean $\delta^{(j)}_t = \mathbf{1}[d^{(j)}_t > T]$. From these, we derive the \emph{delay indicator} $p_t \in \{0,1\}$, defined as $p_t = \mathbf{1}\!\left[\sum_{j} \delta^{(j)}_t \geq f+1\right]$, which the policy consumes as one coordinate of its state. If $f+1$ nodes were unavailable, only $2f$ would remain, motivating the threshold. Geographic latency or transient congestion can also raise $p_t$, so it does not by itself establish a partition or loss of consensus progress. We refer to the procedure that emits $p_t$ as \texttt{NetworkStatus}.

\vspace{2pt}
\noindent
\textbf{\underline{Reward.}}
In our RL framework, the reward function guides protocol selection by encoding the system's performance objectives. We define the reward at time $t$ as
\[
    r = \alpha_{tp} \cdot tp_t - \alpha_l \cdot l_t + \alpha_{ld} \cdot ld_t - \alpha_s \cdot \hat{p}_t - \alpha_{sw} \cdot \chi_t,
\]
where $tp_t$, $l_t$, $ld_t$, $\hat{p}_t$, and $\chi_t$ denote throughput, latency, offered load, delay penalty, and switch indicator, respectively, each scaled by a weight $\alpha$.
$+\alpha_{tp} \cdot tp_t$ encourages higher transaction rates, while $-\alpha_l \cdot l_t$ rewards lower latency. The action-independent load term $\alpha_{ld}\cdot ld_t$ favours neither protocol and keeps rewards comparable across loads.
The delay penalty $\hat{p}_t$ equals $p_t$ when HotStuff is selected and $0$ for FIN, so only the leader-based protocol incurs this penalty. The switch indicator $\chi_t$ is $1$ for a protocol switch and $0$ otherwise, with $\alpha_{sw} \cdot \chi_t$ discouraging excessive switching to maintain stable performance.

\subsection{Instantiation with DQN}
\label{subsec:dqn-training}

We instantiate the learning module at each node using DQN~\cite{mnihPlayingAtariDeep2013a}. The state includes continuous performance measurements, while the action space is discrete, consisting of HotStuff and FIN. DQN approximates the action value for each candidate protocol to guide local proposals.

\vspace{3pt}
\noindent
\textbf{Learning Objective.}
The neural network is trained by minimizing the Huber loss $\ell_H$ of the temporal-difference residual:

{
    \vspace{-8pt}
    \small
    \[
        \mathcal{L}_i(\theta_i) = \mathbb{E}_{s, a, r, s' \sim \mathcal{D}_i} \left[ \ell_H\!\left( r + \gamma \max_{a'} Q(s', a'; \theta_i^-) - Q(s, a; \theta_i) \right) \right]
    \]
    \vspace{-4pt}
}

where $\mathcal{D}_i$ is the experience replay buffer, $\theta_i$ and $\theta_i^-$ are the active and target network parameters, and $\gamma$ balances immediate and future rewards. This objective minimizes the discrepancy between the predicted action value and a target combining the observed reward with the discounted next-state value. The next-state term uses the highest action value predicted by the target network. Thus, \sysname's policy can propose a switch despite its transient latency cost when the expected performance gains over subsequent windows outweigh that cost. We defer full details of our DQN training configuration to Appendix \ref{appendix:sec-dqn}.

\subsection{Metrics Synchronization}
\label{metrics-synchronization}

\begin{algorithm}[t]
    \caption{Metrics Synchronization Utility}
    \label{alg:metrics-sync}
    \tcc{$\widetilde{S}$: coordinate-wise median of multiset $S$; the mean of the two middle values for even $|S|$}

    \small
    \SetKwFunction{FLocal}{LocalMetrics}
    \SetKwFunction{FAggreed}{AgreedMetrics}
    \SetKwProg{Fn}{Function}{}{end}

    \Fn{\FLocal{$metrics$}}{
        $lats, \; tp \gets \{m.latency : m \in metrics\}, \; \text{window throughput}$\;
        $rtt_j \gets$ probe round trip to each peer $j$\;
        \Return{$(\widetilde{lats}, \; tp, \; \{rtt_j\})$}\;
    }

    \BlankLine

    \Fn{\FAggreed{$reports$}}{
        $lats, \; tps \gets$ \{r.latency\}, \{r.throughput\} for $r \in reports$\;
        $d_j \gets \widetilde{\{r.rtt_j\}}$ over reports $r$ observing peer $j$, one value per reporter, kept if at least $f{+}1$ reporters observed $j$\;
        \Return{$(\widetilde{lats}, \; \widetilde{tps}, \; \{d_j\}, \; \textit{NetworkStatus}(\{d_j\}))$}\;
    }

\end{algorithm}

Algorithm~\ref{alg:metrics-sync} aligns learning inputs across replicas without introducing a separate coordination protocol. Each replica summarizes a window $w$ using \texttt{LocalMetrics}, with $tp$ computed as committed bytes divided by window duration, and attaches the report to its next protocol message to each peer. Each per-peer round trip includes the probed peer's signature, authenticating the response but not its freshness in the reporting window. The prober measures the round trip, but the median limits manipulation by faulty probers when that peer coordinate has at least $2f{+}1$ observations. Aggregation requires at least $2f{+}1$ authenticated reports for the same window, with at most one report per replica. Each replica applies \texttt{AgreedMetrics} to the reports it holds, and a switch vote binds the window, rounded aggregate latency and throughput, partition flag, and contributor identities. Thus, a $2f{+}1$ certificate requires matching digests, which exclude peer delays and the delayed-peer fraction. The offered load $ld_t$ is each replica's own client rate and is not aggregated. The agent consumes the sample online, whereas the model is trained offline. \S\ref{sec:analysis} establishes the bound on Byzantine influence.

\section{Analysis}
\label{sec:analysis}

We give a conditional safety argument under the $n=3f{+}1$ Byzantine model and the handoff rules of \S\ref{subsec:base-system}. Certificate holders refuse incumbent decisions above $b_w$ and initialize the target from the prefix and deduplication state through $b_w$. The timing premise requires at least $f{+}1$ honest replicas to receive the certificate before sending any votes or epoch messages supporting incumbent decisions above $b_w$.

\textbf{Agreement.} 
Any two certificates for window $w$ name the same target and boundary. Each requires $2f{+}1$ distinct signers, so their signer sets intersect in at least $f{+}1$ replicas, including an honest one, and that replica would have to sign two different votes, contrary to the one-vote-per-window rule. A replica that receives a certificate rather than assembling it verifies the same $2f{+}1$ signatures. Agreement on the prefix at $b_w$ follows from the incumbent's safety.

\textbf{Total order.} 
Let $L_{\leq b_w}$ be the committed application prefix through $b_w$. FIN supplies agreed payloads in a fixed deterministic order, and HotStuff supplies ordered blocks. Under the stopping condition, at least $f{+}1$ honest replicas never support incumbent decisions above $b_w$, leaving at most $2f$ participants and preventing a $2f{+}1$ quorum. The target's first decision extends $L_{\leq b_w}$. Its agreement and ordering properties preserve a single extension, so induction over successive handoffs preserves global order. Matching local round numbers alone would not establish this property.

\textbf{Request execution.}
Re-proposed requests retain their client identity and sequence number. Because the target inherits the boundary's deduplication state and continues the same execution rule, a previously executed request cannot execute again. This establishes at-most-once execution across the handoff, conditional on that state continuity. Eventual inclusion of an uncommitted request additionally requires continued dissemination and the target's request-inclusion guarantees.

\textbf{Liveness.}
If the incumbent commits through $b_w$, honest replicas receive the certificate, and the prefix remains available for verified catch-up, replicas can complete the handoff. They need not activate simultaneously. With $n{-}f$ honest replicas in the same target instance, progress follows under the target's liveness conditions. Votes and certificates propagate directly among replicas. The boundary lies $k$ windows after the voting window, but dissemination or catch-up may delay activation. Poor policy choices or repeated switching can impair performance without bypassing safety checks. We claim no escape from an incumbent stalled before $b_w$.

\textbf{Byzantine robustness of the metric aggregation.}
A natural concern is whether a Byzantine replica can poison the policy through the metric aggregation. We use the same $n=3f{+}1$ model as in \S\ref{subsec:network-model}, with at most $f$ faulty replicas. A faulty replica may submit arbitrary valid values for any metric coordinate. The median-based aggregation in Algorithm~\ref{alg:metrics-sync} bounds their influence as follows.

\emph{Two-stage robust aggregation.}
Each replica reports median latency and window throughput. Cross-replica aggregation uses coordinate-wise medians. The replica fault bound alone provides no quantitative guarantee against temporal outliers. The Byzantine bound below follows from the honest majority in the cross-replica stage.

\emph{Robustness bound.}
Let $R_w$ be the collection of $m\geq 2f{+}1$ reports, one per replica, aggregated for window $w$ (\S\ref{metrics-synchronization}). Votes with different aggregate digests do not combine. Let $H_w\subseteq R_w$ contain the honest reports. Since at most $f<m/2$ reports are Byzantine, honest reports form a strict majority. Thus, for every coordinate $k$ present in all $m$ reports,
$\min_{x\in H_w} x^{(k)} \leq \widetilde{R_w}^{(k)} \leq \max_{x\in H_w} x^{(k)}$.
A median outside this interval would require at least half the reports to be outside the honest range, contradicting $f<m/2$. Deterministic aggregation gives all honest replicas the same result for identical inputs. This bound applies directly to latency and throughput coordinates, which have one value per report. A per-peer delay coordinate has one observation per reporter and is retained once $f{+}1$ reporters contribute. Its median lies in the honest range when the contributing reporters have a strict honest majority, which $2f{+}1$ observations ensure. The $f{+}1$ retention threshold alone does not ensure this bound. Even where the bound holds, global accuracy is not guaranteed, and Byzantine bias may persist across windows.

\section{Implementation and Evaluation}
\label{sec:eval}

Our \sysname node prototype uses Python for its main routines\footnote{\label{fn:hotswitch}Available at: \url{https://anonymous.4open.science/r/hotswitch}.} and implements PBFT, HotStuff, Dumbo, and FIN from their original papers~\cite{castro1999practical,yin2019hotstuff,guo2020dumbo,duan2023fin}. Its three-layer DQN has two 64-unit hidden layers over the six-dimensional state and is trained offline in a simulator calibrated to testbed latency and throughput.
Switching is restricted to HotStuff and FIN to study adaptation between leader-based partially synchronous and leaderless asynchronous BFT.
We adapted BFTBrain~\cite{wu2025bftbrain}\footnote{\label{fn:bftbrain}Adapted from: \url{https://github.com/JeffersonQin/BFTBrain}.}, a state-of-the-art BFT switching scheme, to the same action space by adding \sysname's FIN implementation while retaining BFTBrain's throughput-based reward. Its decentralized learner trains online from an empty experience buffer each run.
Linux's traffic control utility \texttt{tc}, from \texttt{iproute2}~\cite{tc_utility}, delays outbound packets on each node's network interface to emulate varying delay patterns, including adaptive attacks.
We deploy on CloudLab's Emulab cluster~\cite{Duplyakin+:ATC19} using \texttt{d710} nodes with Intel Xeon E5530 processors (8 cores, 2.4~GHz), 12~GB RAM, and no GPU. All offline training and inference run on CPUs.

\subsection{Evaluation Setting}
Our evaluation asks: \emph{Can \sysname efficiently adapt between leader-based partially synchronous and leaderless asynchronous BFT protocols under dynamic and potentially adversarial network conditions?} Using DQN for selection, we examine switching behavior and overhead and compare latency and throughput with static protocols and BFTBrain. Appendix~\ref{app:eval-more} examines scaling, activation lag, and the policy's network measurements and decisions.

\textbf{Experimental Network.}
Our primary experimental configuration uses $n{=}31$ replicas, 250~B transactions, and $w{=}5$. Smaller configurations ($n{=}4$ and 10) support candidate benchmarks, scale comparisons, and sensitivity tests. Each replica generates requests independently with open-loop Poisson arrivals, and consensus batches the queued requests. We vary the per-replica arrival rate to control \emph{offered load}, defined as the aggregate request arrival rate times transaction size. Offered load and committed payload throughput are reported in KB/s.
The evaluation exercises switching, metric synchronization, and adaptation under controlled adverse conditions. HotStuff's view timer is disabled in the main delay experiments. Exceptions use fixed timers of 4~s for unresponsive replicas (\S\ref{subsec:eval-silent}), 10~s for BFTBrain's HotStuff in the varying-load experiments, and 5~s in the sensitivity experiments.

Latency measures the interval from local request generation to commitment at the originating replica, including queuing and consensus. For the $n{=}31$ results, we report the median across replicas' p50 latency values unless otherwise stated.

\textbf{Network Conditions.} We simulate four network delay conditions
with \texttt{tc}, which emulates the effects of geographical distances and adversarial influences.

$\bullet$ \textit{Safe:}
Without additional delay or node attacks.

$\bullet$ \textit{Leader Delay:}
The attacker delays the HotStuff leader's outbound packets and can track the leader across rounds.
Under FIN, the same delay is applied to one fixed replica.

$\bullet$ \textit{Global Delay:} 
The same additional one-way delay is applied to every replica's outbound packets, emulating a network-wide slowdown.

$\bullet$ \textit{Jitter:}
Each replica's outbound delay is resampled at each committed height in the range 0--10~ms using clipped normal samples, emulating variable network conditions.

Injected delays are one-way delays. Their magnitudes and phase lengths are specified for each experiment.

\begin{figure}
    \centering
    \includegraphics[width=0.98\linewidth]{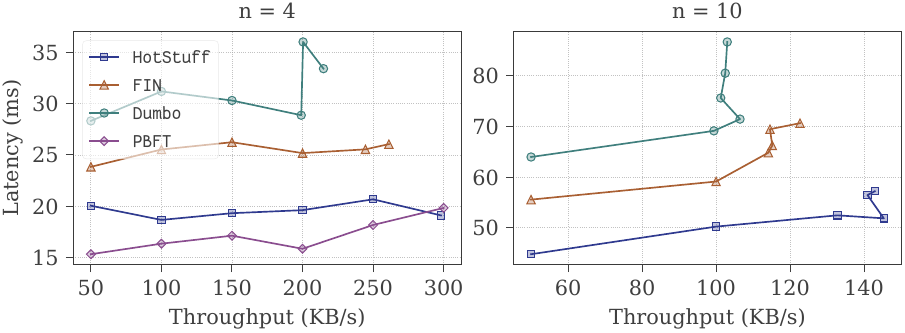}
    \caption{
    Latency vs. throughput of candidate protocols under varying network loads in the safe network condition.
    }
    \label{fig:candidate_lat_tp}
\end{figure}

\subsection{Benchmarking Candidate Protocols}

Fig.~\ref{fig:candidate_lat_tp} shows throughput vs latency of HotStuff, FIN, PBFT, and Dumbo as the batch rate increases from 2 to 12 batches/s, corresponding to offered loads from 50 KB/s to 300 KB/s.  When $n{=}4$, PBFT has the lowest latency (15--20 ms).  All remain unsaturated with no noticeable queuing delay except Dumbo. When $n{=}10$, latency increases for all protocols: HotStuff remains lowest (40--55 ms), FIN ranges from 56 to 71 ms, Dumbo from 65 to 85 ms, and PBFT rises from roughly 179~ms to 2257~ms (not shown). In throughput, HotStuff exceeds 145 KB/s, followed by FIN at about 125 KB/s, Dumbo at 110 KB/s, and PBFT at 43 KB/s. These results motivate our selection of HotStuff and FIN as representative leader-based partially synchronous and leaderless asynchronous protocols, respectively, for integration with our RL agents.

\begin{figure}
    \centering
    \includegraphics[width=0.98\linewidth]{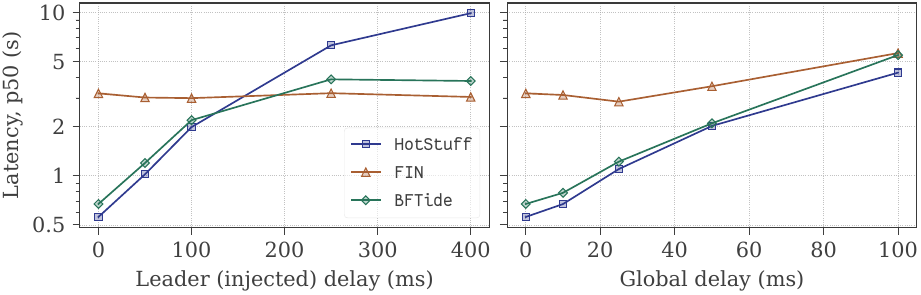}
    \caption{Median latency of HotStuff, FIN, and \sysname under leader delay (left) and global delay (right).}
    \label{fig:hs_performance}
\end{figure}

\begin{figure*}
    \centering
    \begin{subfigure}{0.46\textwidth}
        \centering
        \includegraphics[width=\textwidth]{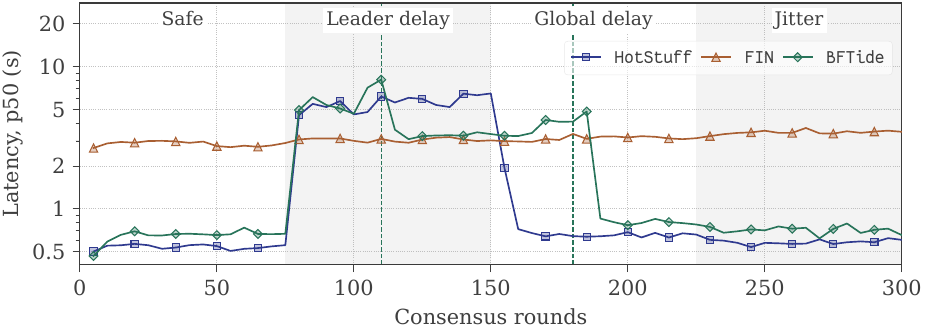}
        \caption{Local Network}
        \label{fig:latency_comparison}
    \end{subfigure}
    \hspace{4pt}
    \begin{subfigure}{0.46\textwidth}
        \centering
        \includegraphics[width=\textwidth]{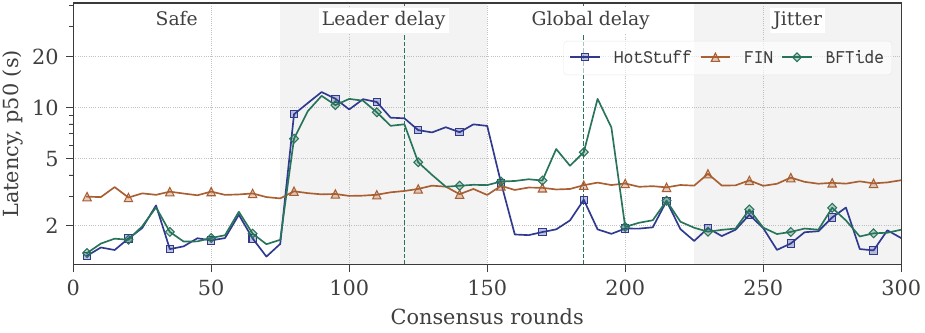} 
        \caption{Four emulated regions (WAN)}
        \label{fig:latency_comparison_wan}
    \end{subfigure}
    \vspace{-0.5em}
    \caption{Latency as the network moves through four conditions ($n{=}31$, 75~KB/s); dashed lines mark \sysname's switches.}
    \label{fig:overall_performance}
\end{figure*}

\subsection{Performance Under Delay Attacks}

Fig.~\ref{fig:hs_performance} compares median latency under leader delay (left) and global delay (right) as the injected one-way delay increases. We fix $n{=}31$ and an offered load of approximately 75.7~KB/s.

In the absence of injected delay, HotStuff outperforms FIN in latency (0.56~s versus 3.20~s). As leader delay increases, however, HotStuff's latency rises rapidly to 9.90~s at 400~ms, whereas FIN remains near 3.0--3.2~s. \sysname initially uses HotStuff, retaining it at 50 and 100~ms delays but dynamically switching to FIN at 250 and 400~ms. Including initial HotStuff operation and the transition, it achieves latencies of 3.90 and 3.81~s, respectively, versus HotStuff's 6.30 and 9.90~s. Thus, \sysname reduces latency under larger leader delays by switching to the better-performing protocol.

Unlike leader delays, global delays, injected uniformly across nodes, increase request latency for both leader-based and leaderless protocols. In \sysname, increased peer delays affect the policy state and may trigger FIN. \sysname retains HotStuff at 10, 25, and 50~ms but switches to FIN at 100~ms, reaching 5.49~s latency---near FIN's 5.64~s but above HotStuff's 4.28~s. Thus, switching does not improve latency. 

\begin{figure}[t]
    \centering
    \includegraphics[width=\linewidth]{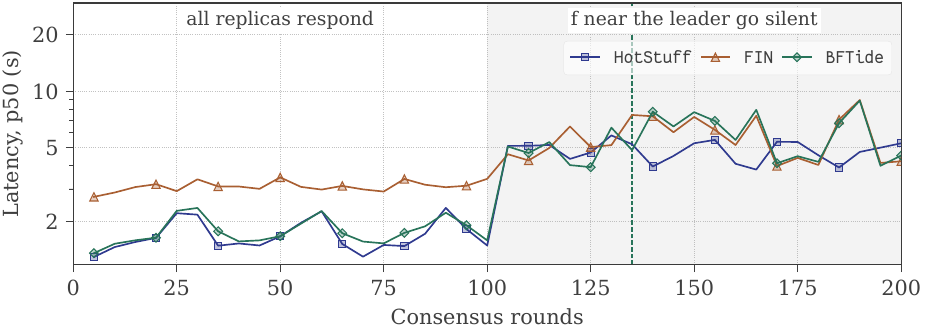}
    \caption{Latency with $f{=}10$ unresponsive replicas ($n{=}31$, emulated WAN); aggregation in Appendix~\ref{app:eval-silent-details}.}
    \label{fig:silent}
\end{figure}

\subsection{Evolving Network Conditions}
\label{subsec:eval-temporal}
\label{subsec:eval-wan}

We assess continuous performance at $n{=}31$ and 75~KB/s offered load over 300 consensus rounds in four 75-round phases: safe, leader delay, global delay, and jitter. Leader delay injects 250~ms at the leader, global delay adds 10~ms at every node, and jitter redraws 0--10~ms per node each round. All injected delays are one-way. Fig.~\ref{fig:overall_performance} compares a LAN and four emulated regions.

\vspace{2pt}
\noindent\textbf{Local network.}
During `Safe' (Fig.~\ref{fig:latency_comparison}), HotStuff and \sysname maintain latency around 0.55--0.67~s, compared with FIN's 2.9~s. Under `Leader Delay', HotStuff's latency rises to 5.6~s while FIN remains stable at 3.1~s.
\sysname switches to FIN at round 110 and tracks its latency. The 35-round response comprises two windows for probes to reflect leader delay, two confirming windows each decided one window after closure, and a three-window lead for certificate dissemination.
After the attack ends at round 150, the leader's probe round trip drops below $T$ within two windows, and \sysname returns to HotStuff at round 180. It tracks HotStuff at 0.81~s during the remaining `Global Delay' phase and 0.71~s during `Jitter', compared with FIN's 3.0--3.7~s. HotStuff sustains approximately 75~KB/s over the run, indicating that the attack primarily affects latency at this load.

\vspace{2pt}
\noindent\textbf{Geographic distribution.}
To assess regional delay effects, we emulate four regions of 7--8 replicas with 20--100~ms one-way delays between regions and LAN delay within each region. Setting $T{=}350$~ms keeps baseline round trips of up to about 200~ms below the threshold.

\begin{table}[t]
    \centering
    \small
    \caption{Median latency under emulated geography at $n{=}31$, with and without a 250~ms leader delay.}
    \label{tab:wan}
    \vspace{-0.75em}
    \begin{tabular}{lcc}
    \toprule
    & \textbf{Regional delays} & \textbf{+ Leader delay} \\
    \midrule
    HotStuff & 1.75 s & 8.52 s \\
    FIN      & 3.34 s & 3.33 s \\
    \sysname & 1.85 s & 4.04 s \\
    \bottomrule
    \end{tabular}
\end{table}

Table~\ref{tab:wan} reports one fixed-condition run per cell, targeting 255 rounds. Regional delays roughly triple HotStuff's median latency (0.56~s to 1.75~s) but barely affect FIN's (3.20~s to 3.34~s), consistent with FIN's cost being dominated by message processing. \sysname retains HotStuff at 1.85~s, including switching-layer overhead. With leader delay, HotStuff reaches 8.52~s while leaderless FIN remains nearly unchanged. \sysname's window latency falls from 8--12~s under HotStuff to 3.3~s after switching to FIN at round 30. Its median latency of 4.04~s includes both periods, as on the LAN. In the phased WAN run (Fig.~\ref{fig:latency_comparison_wan}), switches occur at rounds 120 and 185, within two windows of the LAN timings. After switching back, HotStuff commits requests deferred by FIN's subset selection on five replicas in one window, producing a latency spike that clears within two windows.

\vspace{2pt}
\noindent\textbf{Unresponsive replicas.}
\label{subsec:eval-silent}
When nearby replicas stop responding, a quorum may require replies from more distant replicas. We place $f{=}10$ replicas in the leader's region and drop their outbound traffic from round 100. With a 4~s view timeout, HotStuff advances past silent leaders. \sysname switches to FIN at round 135, and all 21 responsive replicas reach round 200 with no further switches (Fig.~\ref{fig:silent}). The traces show no clear latency benefit. Elevated latency alone need not imply an advantage for FIN, and switching still requires incumbent progress (\S\ref{sec:analysis}).

\begin{figure*}[h]
    \centering
    \begin{subfigure}{0.32\textwidth}
        \centering
        \includegraphics[width=0.97\textwidth]{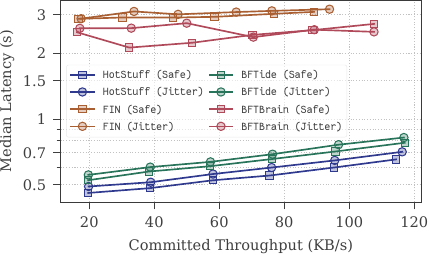}
        \caption{Safe and jitter}
        \label{fig:varying_safe_poisson}
    \end{subfigure}
    \hfill
    \begin{subfigure}{0.32\textwidth}
        \centering
        \includegraphics[width=0.99\textwidth]{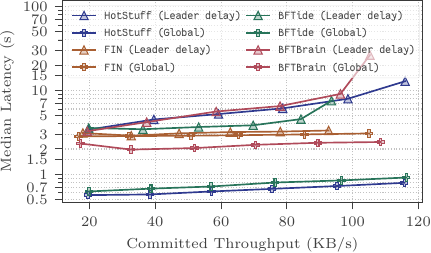}
        \caption{Leader delay and global delay}
        \label{fig:varying_adaptive_global}
    \end{subfigure}
    \hfill
    \begin{subfigure}{0.32\textwidth}
        \centering
        \includegraphics[width=0.99\textwidth]{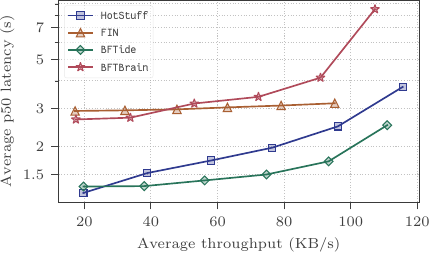} 
        \caption{Average over the four conditions}
        \label{fig:varying_average}
    \end{subfigure}
    \vspace{-0.5em}
    \caption{Latency vs throughput across varying offered loads under different network conditions.
    }
    \label{fig:lat_vs_tp}
\end{figure*}

\subsection{Performance Under Varying Offered Load}

We evaluate \sysname under varying transaction workloads and compare its performance with BFTBrain~\cite{wu2025bftbrain}. We fix $n{=}31$ and vary the offered load from 20 to 120~KB/s in 20~KB/s steps, using $T{=}150$~ms. Each network condition is evaluated separately: safe, 0--10~ms jitter, 250~ms leader delay, and 10~ms global delay. We report the median across available runs for each configuration.

Fig.~\ref{fig:varying_safe_poisson} plots median latency against achieved throughput under the `Safe' and `Jitter' conditions.
Across these conditions, HotStuff and \sysname have the lowest latency, at 0.46--0.71~s and 0.53--0.82~s, respectively, while BFTBrain and FIN remain above 2~s. \sysname retains HotStuff and closely tracks its throughput, with both reaching 114--117~KB/s at the highest offered load. FIN and BFTBrain achieve lower throughput in these runs.

Next, we evaluate performance under leader delay and global delay (Fig.~\ref{fig:varying_adaptive_global}).
Under leader delay, HotStuff's latency grows from 3.41 to 12.87~s as the offered load increases, whereas FIN remains near 2.88--3.32~s. \sysname retains HotStuff at 20~KB/s and switches to FIN at round 30 for loads of 40~KB/s and above. At 120~KB/s, its median latency is 7.53~s, compared with 12.87~s for HotStuff and 26.41~s for BFTBrain. These measurements include the initial operation under HotStuff before switching. Under the 10~ms global delay, \sysname retains HotStuff and maintains 0.62--0.91~s latency, below BFTBrain's 1.97--2.42~s and FIN's 2.83--3.06~s.

Finally, Fig.~\ref{fig:varying_average} averages the four scenario-level statistics at each offered load.
\sysname has the lowest average latency from 40 to 120~KB/s, while HotStuff is slightly lower at 20~KB/s. At 120~KB/s, \sysname achieves 2.51~s compared with BFTBrain's 8.51~s, a reduction of about 70\%. Its average throughput reaches 110.9~KB/s, exceeding BFTBrain's 107.3~KB/s and FIN's 95.2~KB/s, while HotStuff achieves 115.6~KB/s. These averages summarize performance across separate network conditions; they do not imply that \sysname exceeds the best underlying protocol in every condition.

\begin{coloredleftbar}
\noindent\textbf{Why \sysname outperforms BFTBrain.} \textit{Two main factors contribute: (i) embedded switch agreement vs.\ BFTBrain's out-of-band round per switch; (ii) latency-aware reward vs.\ BFTBrain's throughput-only signal, which is what discriminates the two regimes under adaptive attack.
}
\end{coloredleftbar}

\subsection{Protocol Switching Overhead}

The active protocol determines base communication cost. Embedding can share transmissions and decisions, but evidence adds bytes and verification work; control-only values and catch-up can incur further costs. Message count, consensus-decision count, and transition latency are distinct quantities.

\begin{table}
    \centering
    \small
    \caption{Latency overhead without switching and RL runtimes for \sysname at $n{=}31$ and 75~KB/s offered load.}
    \label{tab:overhead}
    \vspace{-0.75em}
    \begin{tabular}{l@{\hspace{0.5em}}c@{\hspace{1em}}|l@{\hspace{0.5em}}}
    \toprule
    \multicolumn{2}{l|}{\textbf{\makecell[l]{Added median latency}}} &
    \textbf{\makecell[l]{RL component runtimes}} \\
    \midrule
    HotStuff & $0.12 \pm 0.01$ s & \makecell[l]{Retraining DQN} $118$ s \\
    FIN      & $0.32 \pm 0.01$ s & \makecell[l]{Inference} $0.18 \pm 0.02$ ms \\
    \bottomrule
    \end{tabular}
\end{table}

We measure idle cost relative to static protocols with the partition predicate disabled and the incumbent always proposed; the PS Layer still probes, reports, and aggregates every window. Table~\ref{tab:overhead} reports three runs each at $n{=}31$ and 75~KB/s. The layer adds $0.12 \pm 0.01$~s to HotStuff's 0.56~s median (21\%) and $0.32 \pm 0.01$~s to FIN's 3.20~s (10\%). Throughput is unchanged for HotStuff and within 5\% for FIN. Per-window CPU work comprises 30 attestation verifications, aggregation, and policy evaluation, shared with consensus. Reports piggyback on incumbent messages, 1500 per replica per run. Separate messages comprise 60 probes and acknowledgements per replica per window. HotStuff's message count rises by 30\% because time-driven request gossip continues throughout the 21\% longer run at the same client rate. Inference costs $0.18 \pm 0.02$~ms. Import and checkpoint loading take about 2.3~s once at replica startup, outside the consensus path. Offline retraining takes 118~s for 40k simulator steps on the same node without involving the replicas.

\subsection{Window Size and Threshold Sensitivity}
\label{sec:sensitivity}

We vary window size $w {\in} \{5,10\}$ rounds and peer-delay threshold $T {\in} \{50,100,150,300\}$~ms RTT (\S\ref{subsec:mdp}) at $n{=}10$ and 24~KB/s offered load, with a 100~ms delay on the leader's outgoing messages. Each configuration uses one 150-round run with a threshold policy. FIN is proposed if aggregate latency exceeds 600~ms or the partition flag is set. HotStuff requires a clear flag and either no aggregated peer RTT above $T$ or, without probe data, latency below 300~ms.

\begin{figure}
    \centering
    \includegraphics[width=\linewidth]{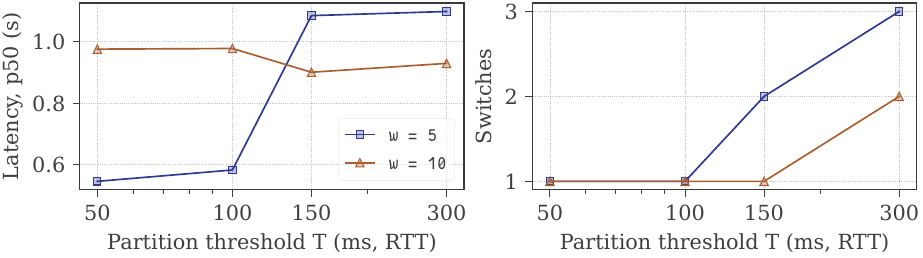}
    \caption{Sensitivity to $w$ and $T$ at $n{=}10$ under a 100~ms leader delay: latency (left) and switch count (right).}
    \label{fig:sensitivity}
\end{figure}

Fig.~\ref{fig:sensitivity} shows that $T$ determines whether the policy settles. With $T$ at or below 100~ms, the leader counts as delayed in every window. \sysname switches to FIN once and stays, with a p50 of 0.55--0.58~s at $w{=}5$. At $T{=}150$~ms, the leader's RTT straddles the threshold, sometimes leaving no peer classified as delayed and allowing the policy to propose a return to HotStuff. The $w{=}5$ run switches twice and reaches 1.09~s, compared with three switches at $T{=}300$~ms. Window size $w$ affects reaction time and switching frequency: the first switch occurs at round 30 for $w{=}5$ and 60 for $w{=}10$. The longer window adds 30 rounds on the delayed leader (0.98~s versus 0.58~s at $T{=}100$~ms), halves re-evaluation frequency, and reduces oscillation at mis-set $T$ (one switch at 150~ms, two at 300~ms).

\begin{coloredleftbar}
\noindent\textbf{Takeaway for sensitivity.} \textit{$T$ must lie below the round-trip excess to be detected. Above it, the indicator detects the delayed peer intermittently and the policy oscillates. $w$ trades reaction time against damping and probe traffic. At $n{=}31$, we fix $w{=}5$ and $T{=}250$~ms for best performance, except for geography ($T{=}350$~ms) and varying load ($T{=}150$~ms).
}
\end{coloredleftbar}

\section{Related Work}
\label{sec:related}

\noindent
\textbf{Dual-path BFT.~} Dual-path BFT solutions combine an optimistic-path protocol for favorable conditions and a pessimistic-path protocol for harsh conditions. They are classified into the \emph{serial} type \cite{blum2019synchronous,gelashvili2021prepared,gelashvili2022jolteon} and the \emph{parallel} type \cite{neu2021ebb,crain2021red,blum2023abraxas,dai2023parbft}. Blum et al.~\cite{blum2019synchronous} introduce a randomized BFT scheme with a fast protocol for synchronous settings and an asynchronous fallback to maintain safety as the network degrades. Gelashvili et al. \cite{gelashvili2021prepared} implement asynchronous fallback through view change for DiemBFT \cite{baudet2019state}, a production version of HotStuff. The same authors propose Jolteon and Ditto~\cite{gelashvili2022jolteon} to combine fast paths with asynchronous fallback in a serial manner with optimized efficiency.
Particularly, Ditto leverages the asynchronous fallback mechanism to guarantee liveness under network partitions.

On the other hand, the parallel-type schemes, including Ebb-and-Flow~\cite{neu2021ebb}, Abraxas \cite{blum2023abraxas}, and ParBFT \cite{dai2023parbft}, run two BFT protocols in parallel: one ``always-live'' asynchronous protocol for accepting client transactions when the network is unstable/partitioned and one fast protocol for confirming transactions when the network is in good condition.

Both types require careful tuning of path-transitioning triggers \cite{blum2019synchronous,gelashvili2021prepared,gelashvili2022jolteon}, unless they let clients determine finality (which path to canonize) based on their synchrony and fault threshold beliefs \cite{neu2021ebb}. In comparison, \sysname enables system nodes to leverage learning-based network intelligence, making informed, proactive decisions that adapt to nuanced network conditions.

\vspace{2pt}
\noindent
\textbf{Multi-protocol Switching.~}
\texttt{Abstract}  \cite{guerraoui2010next} facilitates efficient aborting and switching of BFT protocols within a unified composition framework, while introducing two custom PBFT-style protocols, AZyzzyva and Aliph, as candidates. ADAPT \cite{bahsoun2015making} is an adaptive abortable BFT system that decides ``when to launch which protocol'' based on a quality control module that observes protocol performance amid network condition changes.

The most relevant work to ours is BFTBrain \cite{wu2025bftbrain}, a BFT protocol adaptation system for synchronous and partially synchronous conditions. BFTBrain is the first scheme to use RL to help nodes dynamically select the optimal protocol at runtime. It, however, does not cater to asynchronous conditions or the case where an adaptive adversary targets the leader. BFTBrain's reward prioritizes throughput over latency, which can limit time-sensitive SMR systems. BFTBrain also requires nodes to agree on the new protocol and share local metrics via a separate consensus round. Instead, our design embeds protocol switching logic directly into the base consensus process, reducing communication overhead caused by extra messaging. In our comparison, we extend BFTBrain's action space with FIN to ensure the baseline is not disadvantaged solely due to the protocol pool. \sysname differs in its latency-aware state representation, its reward signals, and its in-band switch mechanism between leader-based and leaderless protocols.

Besides protocol switching, complementary work adapts BFT configurations. AWARE~\cite{berger2020aware} tunes voting weights and leader placement, while Mercury~\cite{berger2024chasing} uses dual resilience thresholds to form smaller quorums under favorable conditions. OptiLog~\cite{gogada2026optilog} assigns replica roles using shared measurement logs and fault monitoring, whereas Beware~\cite{chotkan2026robust} combines latency-report sanitization with robust voting-weight optimization. These approaches also rely on consensus-coordinated adaptation. \sysname instead uses learning-guided switching between protocols; inner-protocol parameter adjustment is left for future work.

\section{Conclusion}
We presented \sysname, an architecture enabling BFT-SMR systems to switch to suitable protocols as network conditions evolve.
\sysname features a lightweight switching mechanism between leader-based partially synchronous and leaderless asynchronous protocols without interrupting SMR operation. Switching remains subject to incumbent progress.
It adapts to its environment, including adversarial delay attacks, using an offline-trained reinforcement learning policy that processes performance metrics to propose protocols.
\sysname outperforms static BFT protocols and the state-of-the-art adaptation scheme in long-term latency and throughput under varied network delays, demonstrating its ability to enhance efficiency in fault-tolerant systems.

\bibliographystyle{IEEEtran}
\bibliography{references/reference,references/HotSwitch, references/class-consensus,references/additional}

\appendix
\section{Further Discussion and Limitations}
\label{sec:discussion}

\textbf{Metric robustness.}The median aggregation in \S\ref{metrics-synchronization} keeps each metric coordinate with at least $2f{+}1$ observations within the range of the included honest reports under the assumptions of \S\ref{sec:analysis}. The median bounds Byzantine influence, but repeated bias within the honest range may still affect the quality of protocol proposals. 

\textbf{Protocol pool.} The system intentionally restricts the protocol pool to HotStuff and FIN to evaluate switching between leader-based partially synchronous BFT and leaderless asynchronous BFT. Extending \sysname to additional protocols, including DAG-based designs, requires implementing the same switch-certificate interface and validating its interaction with each protocol's commit rule. 

\textbf{On scaling.} The primary $n{=}31$ evaluation, supplemented by $n{=}4$ and $10$ experiments, exercises switching, metric synchronization, and delay-regime adaptation under controlled adverse conditions. It provides a limited scale comparison, with timeout settings fixed within each experiment.

\textbf{On periodic offline retraining.} Changes in replica count $N$ or geographic placement can alter protocol performance relative to the policy's training conditions. Periodic offline retraining on representative measurements from the updated deployment may mitigate this mismatch; its effectiveness across deployments remains to be evaluated.

\begin{figure}
    \centering
    \begin{subfigure}{0.278\textwidth}
        \centering
        \includegraphics[height=1.05in]{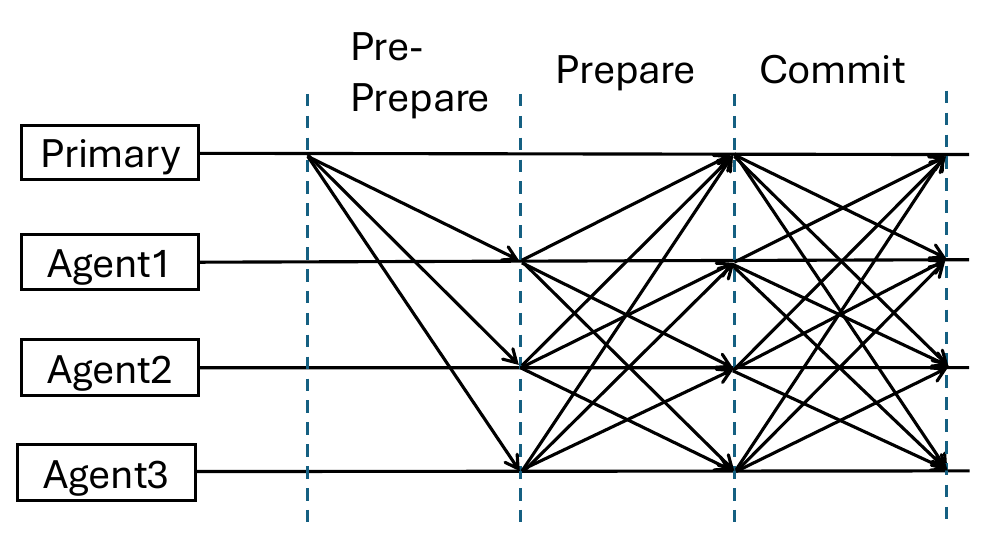}
        \caption{PBFT (main protocol) \cite{castro1999practical}}
    \label{fig:pbft}
    \end{subfigure}
    \begin{subfigure}{0.37\textwidth}
        \centering
        \includegraphics[height=1.05in]{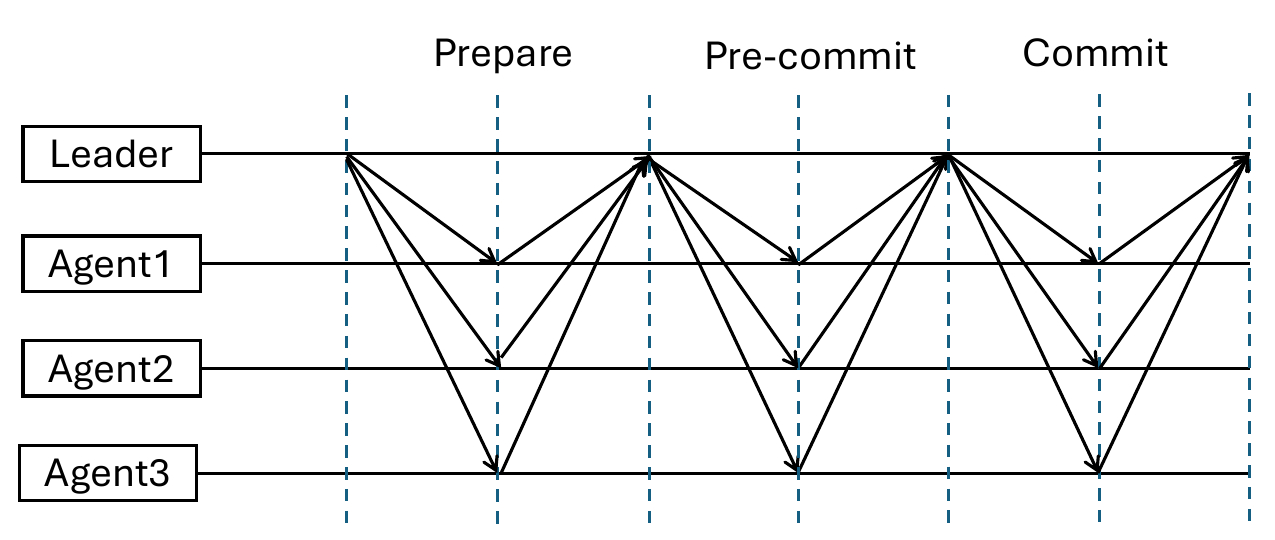}
        \caption{HotStuff (main protocol) \cite{yin2019hotstuff}}
    \label{fig:hotstuff}
    \end{subfigure}
    \begin{subfigure}{0.325\textwidth}
        \centering
        \includegraphics[height=1.04in]{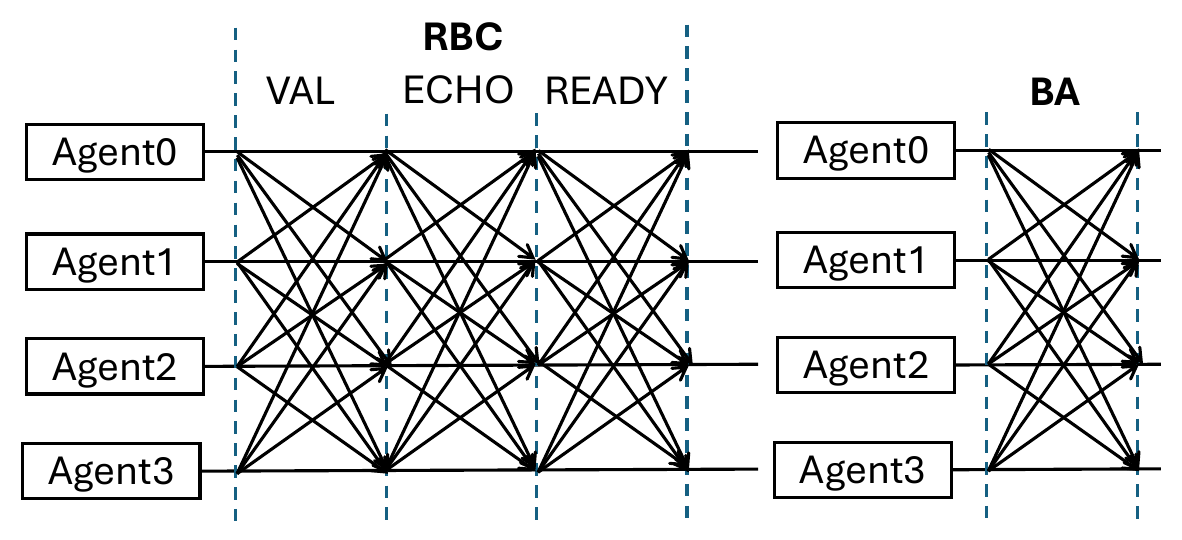}
        \caption{ACS (general form) \cite{ben1993asynchronous}}
    \label{fig:acs}
    \end{subfigure}
    \vspace{-5pt}
    \caption{Comparing BFT protocols in one consensus round (client request handling is not shown).}
    \label{fig:bft-protocols}
\end{figure}

\section{Schematic of Representative BFT Protocols}
\label{app:bft-protocols}

Figure \ref{fig:bft-protocols} shows the main protocol workflow of three representative BFT protocols used in our system (corresponding to the description in \S\ref{subsec:background-bft-protocols}).

\section{DQN Architecture and Training}
\label{appendix:sec-dqn}

This section describes \sysname's DQN architecture, simulator calibration, and offline training configuration.

\subsection{Network Architecture}
We approximate the Q-function $Q(s,a;\theta)$ with a feedforward network in PyTorch~\cite{paszke2019pytorch}. The six state variables of \S\ref{subsec:mdp} enter the network in the following order: latency, throughput, delay indicator $p_t$, offered load, incumbent protocol, and delayed-peer fraction $d_t$. The two outputs estimate the Q-values for HotStuff and FIN, respectively.

The network has two hidden layers of 64 units with ReLU activations and 4{,}738 parameters:
\begin{lstlisting}[language=Python]
Sequential(
  (0): Linear(in_features=6, out_features=64, bias=True)
  (1): ReLU()
  (2): Linear(in_features=64, out_features=64, bias=True)
  (3): ReLU()
  (4): Linear(in_features=64, out_features=2, bias=True)
)
\end{lstlisting}

\subsection{Offline Training Environment}
Training uses a simulator calibrated from the $n{=}31$ testbed at 75~KB/s. The simulator stores baseline latency and the fraction of offered load achieved as throughput for each protocol and network condition. The conditions comprise an undelayed network, leader delays of 50, 100, 200, 250, and 400~ms, global delays of 10, 25, 50, and 100~ms, and a modelled partition. All delay values are one-way. These calibration values are measured except for the 200~ms leader-delay case, interpolated from the measured sweep, and the partition case.

Latency, throughput, and offered load are min-max normalized using ranges of 0.4--8~s, 0--150~KB/s, and 2--12~KB/s per replica, respectively, with values clipped to $[0,1]$. The two network indicators lie in $[0,1]$, and the incumbent is encoded as its protocol index. Training and inference share the same normalization routine.

The simulator assigns a nominal delayed-peer fraction of $1/30$ to leader delays of at least 200~ms, $12/30$ to the partition case, and zero otherwise. These are modelling assumptions for $T{=}250$~ms. In conditions with no nominally delayed peers, one peer is marked delayed with probability 0.05 to represent occasional slow probes. The indicator $p_t$ is set when at least $f{+}1$ peers are delayed.

Each episode spans 64 windows. The initial incumbent is sampled uniformly, as is the offered load from 1.22, 2.44, or 4.88~KB/s per replica. Conditions are sampled with probability 0.25 for the undelayed case and 0.05--0.13 for each remaining case. At each window, the condition is resampled with probability 0.10, and the load moves to an adjacent level with probability 0.10, bounded by the available levels. Latency and throughput are perturbed by multiplicative Gaussian noise with standard deviation 0.08. The reward follows \S\ref{subsec:mdp}, with weights $\alpha_{tp}{=}1$, $\alpha_l{=}1$, $\alpha_{ld}{=}0.5$, $\alpha_s{=}1$, and $\alpha_{sw}{=}0.5$, and is clipped to $[-2,2]$ during training.

\begin{table}[t]
    \centering
    \small
    \caption{DQN training hyperparameters.}
    \label{tab:dqn-hparams}
    \begin{tabular}{ll}
    \toprule
    Environment steps & 40{,}000 \\
    Episode horizon & 64 windows \\
    Discount $\gamma$ & 0.99 \\
    Learning rate (Adam) & $10^{-3}$ \\
    Batch size & 64 \\
    Replay buffer & 10{,}000 transitions \\
    Target network update & every 200 steps \\
    Exploration $\epsilon$ & $1.0 \to 0.05$ \\
    Exploration decay constant & 3{,}000 steps \\
    Loss & Huber (smooth $\ell_1$) \\
    Simulator noise & multiplicative, $\sigma{=}0.08$ \\
    \bottomrule
    \end{tabular}
\end{table}

\subsection{Offline Training Configurations and Combined Workflow with Inference}

\begin{figure}
    \centering
    \includegraphics[width=0.75\linewidth]{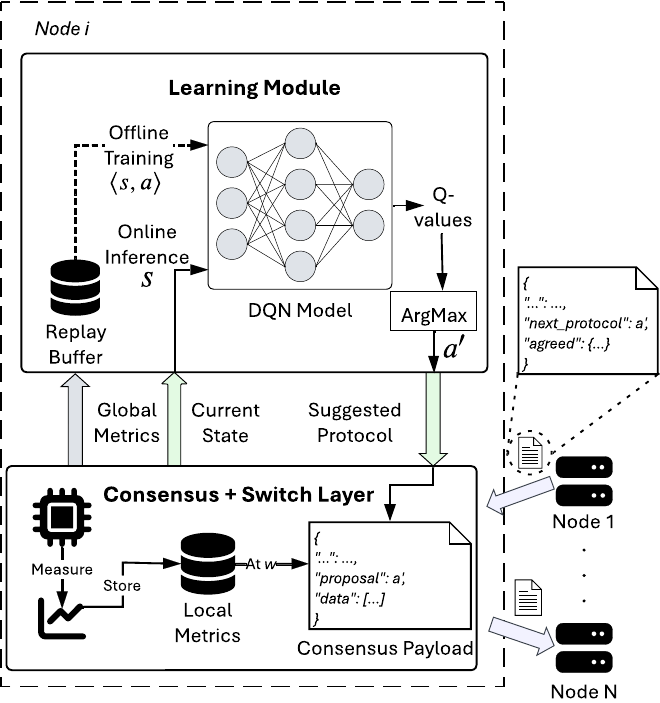}
    \caption{Combined workflow of training and inference of \sysname's RL mechanism.}
    \label{fig:system}
\end{figure}

Table~\ref{tab:dqn-hparams} lists the training hyperparameters. 
We minimise the Huber loss of \S\ref{subsec:dqn-training} using Adam~\cite{kingma2015adam}, uniformly sampled experience replay, and a target network updated by periodically copying the active network's parameters. Exploration follows an exponentially decaying $\epsilon$-greedy schedule. Training for 40{,}000 environment steps with seed 42 takes 118~s. In the inference phase, replicas load the trained checkpoint and propose the protocol with the highest Q-value, keeping model parameters fixed during execution. Inference takes $0.18$~ms (Table~\ref{tab:overhead}). 
Figure \ref{fig:system} shows the combined workflow of training and inference in \sysname's framework.

The varying-load comparison uses an earlier checkpoint calibrated with only the 10~ms global-delay case. Subsequent calibration added the 25--100~ms global-delay cases and the interpolated 200~ms leader-delay case described above.

\section{Additional Evaluation}
\label{app:eval-more}

The experiments in this section extend the evaluation of \S\ref{sec:eval} to the cluster size, to the moment of a switch itself, and to what the switching layer measures and decides. We also give measurement details for the unresponsive-replica experiment.

\subsection{Scaling the Cluster}
\label{subsec:eval-scale}

\begin{figure}
    \centering
    \includegraphics[width=\linewidth]{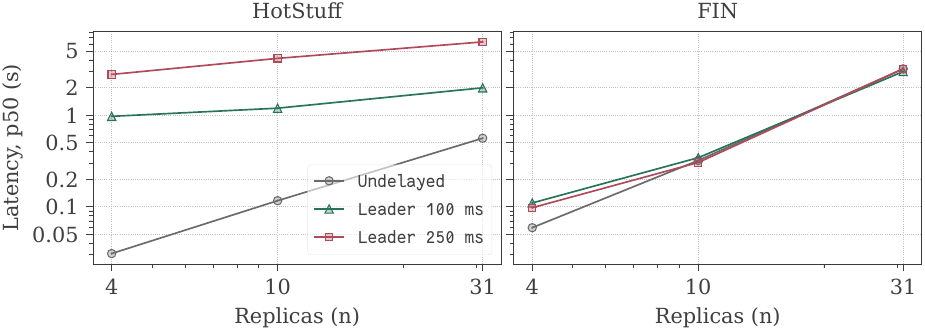}
    \caption{Latency versus replica count under leader delay, for HotStuff (left) and FIN (right).}
    \label{fig:scale}
\end{figure}

Fig.~\ref{fig:scale} repeats the undelayed, 100~ms and 250~ms leader-delay conditions at $n{=}4$, 10 and 31. Both protocols slow with $n$, but not at the same rate. Undelayed HotStuff goes from 0.03~s to 0.12~s to 0.56~s, a few round trips whose cost grows with the leader's fan-out. FIN goes from 0.06~s to 0.32~s to 3.20~s. Its asynchronous common subset uses $n$ parallel reliable broadcasts and one MVBA. The consequence is that the crossover moves. At $n{=}4$ and $n{=}10$ a 100~ms leader delay is already enough for FIN to win (0.97~s and 1.19~s for HotStuff against 0.11~s and 0.34~s for FIN). At $n{=}31$ HotStuff still wins at 100~ms (1.99~s against 2.99~s) and FIN wins only beyond about 150~ms (6.30~s against 3.21~s at 250~ms). FIN's own latency is flat in the delay: a delayed leader is one delayed proposer among $n$, and the subset does not wait for it. This is why the remaining experiments run at $n{=}31$ with a 250~ms leader delay, and why their latencies are in seconds whereas the small-$n$ benchmark of Fig.~\ref{fig:candidate_lat_tp} is in milliseconds.

\subsection{Unresponsive Replica Measurements}
\label{app:eval-silent-details}

Fig.~\ref{fig:silent} reports a single 200-round run for each protocol in the emulated geography described in \S\ref{subsec:eval-silent}. The retained HotStuff and FIN traces show node 1's window-level p50 values, whereas the \sysname trace shows the median across replicas with fresh latency samples in each window. These traces characterize switching behavior rather than compare identically aggregated cluster latencies.

HotStuff records 1158 view timeouts across replicas. The mean of its plotted window p50 values rises from 1.7~s before the faults to 4.8~s during the fault period. FIN and \sysname average 5.7~s over their plotted fault-period window p50 values. FIN's common subset must obtain $n{-}f$ proposals from the remaining responsive replicas. \sysname treats the unresponsive replicas as delayed and switches to FIN at round 135. All 21 responsive replicas reach round 200 in each protocol's run; the ten unresponsive replicas stop at round 100. No further switches occur in the \sysname run.

\subsection{Protocol Activation across Replicas}
\label{subsec:eval-skew}

\begin{figure}
    \centering
    \includegraphics[width=0.7\linewidth]{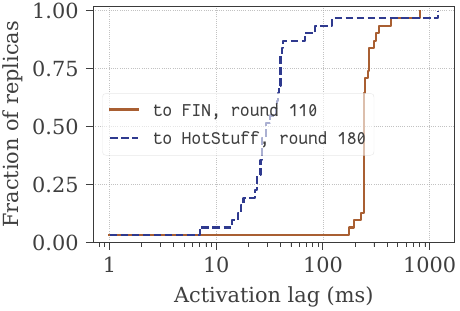}
    \caption{Activation-lag CDF relative to the first replica, for both switch directions ($n{=}31$).}
    \label{fig:skew}
\end{figure}

Message delays and catch-up can leave replicas at different stages of a switch, even when they agree on its target and committed prefix. We measure activation lag to quantify how far apart replicas enter a switch. Fig.~\ref{fig:skew} reports the lag between recorded switch events relative to the first replica for each transition in one run of the phased scenario. For the switch to FIN at round 110, all 31 replicas record the switch within approximately 0.81~s, with a median lag of 243~ms. For the switch back to HotStuff at round 180, 30 replicas record it within 0.3~s (median lag 29~ms), while the remaining replica records it after 1.20~s. The logged timestamps mark entry into the switch routine, before target initialization.

Preserving the committed log despite these timing differences depends on the handoff requirements in \S\ref{subsec:base-system}: the incumbent cannot extend the agreed prefix beyond its terminal height, and the target must extend that same prefix. A lagging replica verifies the switch certificate and obtains and executes any missing committed data before activating the target. These requirements permit different activation times while preserving a common committed history. Agreement on the terminal height alone is insufficient.

\subsection{Network Measurements and Policy Decisions}
\label{app:eval-extra}

Protocol proposals depend on both measured network conditions and the policy's interpretation of them. We inspect the aggregated delay signals to check whether they expose the injected conditions, and the learned decision map to examine how latency, delayed-peer fraction, and the incumbent affect proposals.

\begin{figure}
    \centering
    \includegraphics[width=\linewidth]{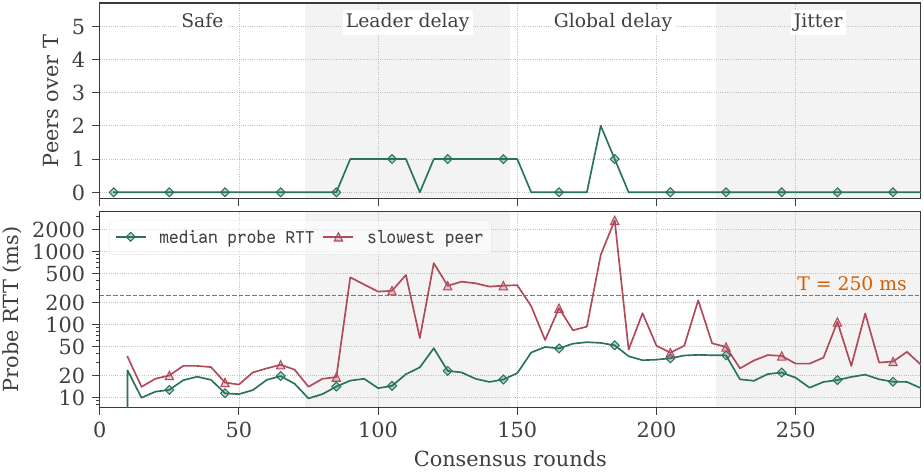}
    \caption{Aggregated delayed-peer count (top) and probe RTT (bottom) during a phased run.}
    \label{fig:trigger}
\end{figure}

\begin{figure}
    \centering
    \includegraphics[width=\linewidth]{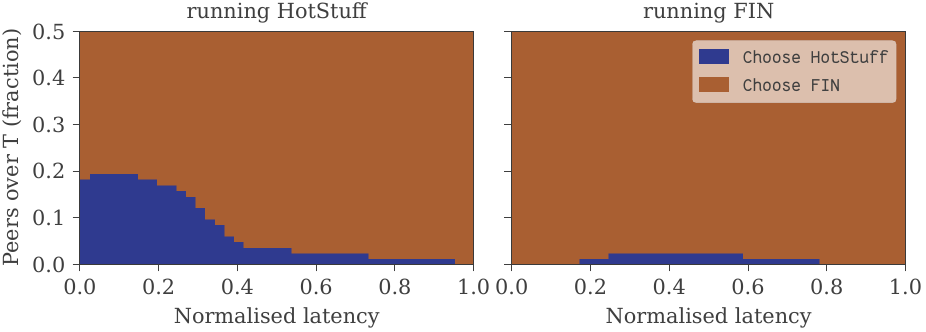}
    \caption{Policy proposals over latency and delayed-peer fraction, with HotStuff (left) or FIN (right) active.}
    \label{fig:decision_map}
\end{figure}

Fig.~\ref{fig:trigger} shows the aggregated delay signals across the phased run. No peer exceeds $T$ during the safe and jitter phases. During most of the leader-delay phase, one peer exceeds $T$ while the median probe RTT remains below it. The delayed-peer count briefly returns to zero. Global delay raises the median RTT, with a transient threshold crossing involving two peers near round 185 during switch-back. Tracking individual peers therefore reveals leader delay even when the median RTT does not cross $T$. These traces characterize the response to network delays. Resistance to falsified reports is governed by the aggregation bound in \S\ref{sec:analysis}: under the stated fault assumptions, a median coordinate with at least $2f{+}1$ observations remains within the range of included honest reports, although bias within that range can still affect proposals.

Fig.~\ref{fig:decision_map} examines how the learned policy combines latency, delayed-peer fraction, and the incumbent protocol at the evaluated load. With HotStuff active, the policy retains it over a larger range of delayed-peer fractions at low latency. This range narrows as latency increases. Above a delayed-peer fraction of approximately 0.2, it selects FIN throughout the plotted latency range. With FIN active, it proposes HotStuff only in a narrow region near zero delayed peers and intermediate normalized latencies. The different regions show that the same latency can produce different proposals depending on the delayed-peer fraction and the incumbent.

\end{document}